\documentclass{article}
\usepackage{amssymb}
\usepackage{graphicx}
\usepackage{amsmath}

\input{tcilatex}
\begin{document}

\begin{center}
{\LARGE Inflationary Cosmology as a Probe of Primordial Quantum Mechanics}

\bigskip

Antony Valentini

\bigskip

\textit{Centre de Physique Th\'{e}orique, Campus de Luminy,}

\textit{Case 907, 13288 Marseille cedex 9, France}

and

\textit{Theoretical Physics Group, Blackett Laboratory, Imperial College
London, Prince Consort Road, London SW7 2AZ, United Kingdom.\footnote{%
Present address.}}

email: a.valentini@imperial.ac.uk

\bigskip
\end{center}

We show that inflationary cosmology may be used to test the statistical
predictions of quantum theory at very short distances and at very early
times. Hidden-variables theories, such as the pilot-wave theory of de
Broglie and Bohm, allow the existence of vacuum states with non-standard
field fluctuations (`quantum nonequilibrium'). We show that inflationary
expansion can transfer microscopic nonequilibrium to macroscopic scales,
resulting in anomalous power spectra for the cosmic microwave background.
The conclusions depend only weakly on the details of the de Broglie-Bohm
dynamics. We discuss, in particular, the nonequilibrium breaking of scale
invariance for the primordial (scalar) power spectrum. We also show how
nonequilibrium can generate primordial perturbations with non-random phases
and inter-mode correlations (primordial non-Gaussianity). We address the
possibility of a low-power anomaly at large angular scales, and show how it
might arise from a nonequilibrium suppression of quantum noise. Recent
observations are used to set an approximate bound on violations of quantum
theory in the early universe.

\bigskip

\bigskip

1 Introduction

2 Quantum equilibrium and quantum nonequilibrium

3 Pilot-wave field theory on expanding space

4 Quantum nonequilibrium in the very early universe

5 Measuring primordial quantum fluctuations

6 Time evolution of nonequilibrium vacua

7 Nonequilibrium power spectrum

8 General remarks

9 Bound on primordial quantum nonequilibrium

10 Possible low power anomaly at small $l$

11 Non-random phases and inter-mode correlations

12 Conclusion

\bigskip

\bigskip

\bigskip

\bigskip

\bigskip

\bigskip

\bigskip

\bigskip

\bigskip

\bigskip

\bigskip

\bigskip

\bigskip

\bigskip

\bigskip

\bigskip

\bigskip\bigskip

\bigskip

\section{Introduction}

According to inflationary cosmology \cite{LL00}, the early universe
underwent a period of exponential expansion, during which microscopic
quantum fluctuations were stretched to macroscopic scales. The resulting
(classical) primordial perturbations seem to be of the form required to
explain the observed temperature anisotropy in the cosmic microwave
background (CMB), and are widely believed to have seeded the formation of
large-scale structure generally. In this scenario, precision CMB
measurements today can provide information about --- and tests of ---
microscopic physics in the very early universe. For this reason, many
workers have turned to inflationary CMB predictions in the hope that these
will provide a `cosmic microscope' with which to probe high-energy physics
at very short distances and at very early times. However, if the primordial
perturbations do indeed have a quantum origin, then inflationary CMB
predictions will also be sensitive to the structure of quantum theory
itself, as well as to that of high-energy physics. Therefore, inflationary
cosmology and CMB measurements may equally be used to probe possible
deformations of quantum theory at very short distances and at very early
times.

In a typical inflationary scenario, at very early times the cosmological
scale factor $a(t)$ undergoes a period of approximately exponential growth, $%
a\propto e^{Ht}$ with $H\approx \mathrm{const}.$. During inflation, field
perturbation modes have physical wavelengths $\lambda _{\mathrm{phys}%
}=a(t)\lambda \propto e^{Ht}$. (As usual, $\lambda =2\pi /k$ is the
wavelength today --- the `comoving wavelength' --- and we set the scale
factor today to be $a_{0}=1$.) A mode `exits' the Hubble radius $H^{-1}$
when $\lambda _{\mathrm{phys}}\gtrsim H^{-1}$, at a time $t_{\mathrm{exit}%
}=t_{\mathrm{exit}}(k)$ (which can be defined by $2\pi a(t_{\mathrm{exit}%
})/k\sim H^{-1}$ or by $a(t_{\mathrm{exit}})/k\sim H^{-1}$). Soon after $t_{%
\mathrm{exit}}(k)$, the perturbation `freezes' and becomes part of the
primordial spectrum. After inflation ends, physical wavelengths $\lambda _{%
\mathrm{phys}}\propto a$ grow more slowly than the Hubble radius $%
H^{-1}\equiv a/\dot{a}\propto t$ (where $a\propto t^{1/2}$ or $t^{2/3}$, for
radiation-dominated or matter-dominated expansion respectively). Mode
`re-entry' occurs at a time $t_{\mathrm{enter}}(k)$ when $\lambda _{\mathrm{%
phys}}\lesssim H^{-1}$, after which the (formerly-frozen) perturbations
begin to grow, eventually giving rise to anisotropies in the CMB and to
large-scale structure \cite{Pad93}.

While there are many uncertainties surrounding the details of inflationary
cosmology, there is a broad consensus that the formation of (frozen)
primordial perturbations takes place when the corresponding physical
wavelengths $\lambda _{\mathrm{phys}}\gtrsim H^{-1}$ are truly microscopic.
Further, because of the huge expansion during the inflationary phase, the
relevant modes will have had very short physical wavelengths, $\lambda _{%
\mathrm{phys}}<<H^{-1}$, at the onset of inflation (where the shorter the
wavelength, the later the time at which the mode exits the Hubble radius
during the inflationary phase). Indeed, it appears that even modes with
initial $\lambda _{\mathrm{phys}}\lesssim l_{\mathrm{P}}$, where $l_{\mathrm{%
P}}\approx 10^{-33}\ \mathrm{cm}$ is the Planck length, may contribute to
the primordial spectrum \cite{TransP}. Clearly, if inflation did indeed
occur, then precision measurements of the CMB (and of large-scale structure
generally) can probe physics at very early times and at very short distances
(possibly even at distances $\lesssim l_{P}$, to the extent that this might
be meaningful).

A number of possible deformations of high-energy physics have been
considered in an inflationary context. These include: (a) modified
dispersion relations (which may be introduced ad hoc \cite{TransP,MDRs1}, or
which may be motivated by quantum-gravitational deformations of Lorentz
invariance \cite{KG01} or by quantum cosmology \cite{HW04}); (b) an
ultra-violet cutoff coming from a fundamental length associated with
deformed uncertainty relations (possibly associated with quantum gravity or
string theory) \cite{UV}; (c) short-distance non-commutative geometry \cite%
{NCG}. Some authors consider that changes in physics at very high energies
may have an effective description in terms of different choices of quantum
vacuum \cite{Dan02} (for a review, see ref. \cite{GSS05}). Excited,
non-vacuum states have also been considered \cite{nonV}. However, while it
is generally agreed that inflationary primordial perturbations have a
quantum origin, effects on the CMB arising from possible deformations of
quantum theory itself are not usually considered. (By `quantum theory' we
mean, essentially, the representation of physical states in Hilbert space,
with unitary evolution, and with probabilities given by the Born rule.) An
exception is Perez \textit{et al}. \cite{PSS06}, who discuss how predictions
for the CMB could be affected by a hypothetical dynamical collapse of the
wave function, a proposal that is motivated by the quantum measurement
problem (which seems especially severe in a cosmological setting).

Despite the widespread reluctance to consider deformations of quantum theory
itself, there is in fact no good scientific reason for believing that the
structure of standard quantum theory is `final', or that the predictions of
quantum theory will continue to hold under all conditions. The following
arguments are often presented as evidence for the finality of quantum
theory: that it provides a universal framework applicable to all systems
independently of their composition (electrons, fields, atoms etc.); that it
is based on simple, elegant axioms; that it provides the basis for powerful
new technologies; and, of course, that in all cases so far it agrees with
experiment. However, arguments similar to these could have been made in the
eighteenth and nineteenth centuries concerning the status of Newtonian
mechanics: at that time, Newtonian mechanics seemed to provide a universal
framework applicable to all systems independently of their composition
(rocks, fluids, planets etc.); it was based on simple, elegant axioms
(Newton's three laws of motion); it provided the basis for powerful new
technologies; and it agreed with all experiments performed to date. And yet,
we now know that Newtonian mechanics is in fact merely approximate and
emergent, arising from a classical and low-energy limit of relativistic
quantum field theory. Of course, that Newtonian mechanics proved to be
approximate and emergent does not imply that quantum theory will necessarily
turn out likewise. However, the case of Newtonian mechanics does suggest
that the above (frequently-cited) arguments for the finality of a physical
theory are not reliable.

The ultimate test of the domain of validity of a scientific theory is, of
course, experiment. No matter how well a theory has been tested in the past,
it will always be subject to possible modification in the future, in
hitherto untested regimes. Therefore, in order to expand our knowledge of
the domain of validity of any given theory, it is necessary to subject it to
ever more stringent tests in ever more extreme conditions. To accomplish
this, it is helpful to have a `foil' against which to test the theory in
question --- that is, to have a model reducing to the given theory only in
some limit.

In the case of quantum theory, a number of alternatives or foils might be
considered. Models with a nonlinear evolution or with a dynamical collapse
of the wave function have, for example, been subjected to considerable
experimental scrutiny. In this paper, we focus on a different possibility:
that of nonequilibrium hidden variables \cite%
{AV91a,AV91b,AV92,AV96,AV01,AV02a,AV02b,AVPr02,AVSig,AVBHs,VW05,PV06,AV07,AV09}%
.

A deterministic hidden-variables theory, such as the pilot-wave theory of de
Broglie \cite{deB28,BV} and Bohm \cite{B52}, agrees with quantum theory only
in the limit in which the hidden parameters have a particular `quantum
equilibrium' distribution \cite{AV91a,AV91b,AV92,AV02a,PV06,AV09}. A foil
against which to test quantum theory may then be obtained from such a theory
by allowing the hidden variables to have a non-standard or `quantum
nonequilibrium' distribution, resulting in statistical predictions that
deviate from those of quantum theory \cite{AV07}.\footnote{%
Note the clear distinction from the foils based on local hidden-variables
models \cite{Bell64} or on a particular restricted class of nonlocal models 
\cite{Leg03}: such models disagree with the quantum predictions for \textit{%
any} distribution (equilibrium or otherwise) of the hidden variables.} Such
possible corrections to quantum theory will be explored here, in the context
of inflationary cosmology, where it will be shown how CMB observations may
be used to set bounds on the presence of quantum nonequilibrium at very
short distances and very early times.

If anomalies are observed in the CMB, one must of course ask if they are
caused by corrections to quantum theory or by some other effect. (For
example, the quantum state during the inflationary phase might differ
significantly from the standard Bunch-Davies vacuum \cite{nonV}.) A similar
issue arises for other proposed corrections to standard physics in the early
universe. Ideally, one would like to find a unique signature that could not
be predicted by any quantum state compatible with inflation. In practice,
one would at least require a quantitative prediction of a deviation from
standard results. The present paper focusses on showing that early quantum
nonequilibrium -- for a given (standard) quantum state -- could have
observable consequences for the CMB. We also sketch two scenarios that would
lead to a specific prediction: for example, deviations for wavelengths
larger than a certain (predicted) infra-red cutoff. But the full development
of these scenarios, and the extraction of precise quantitative predictions
from them, is left for future work.

In section 2, we review the notion of quantum nonequilibrium, in de
Broglie-Bohm theory and in general (deterministic) hidden-variables
theories, and we provide motivation for why quantum nonequilibrium might
exist in the very early universe. In section 3, we develop pilot-wave field
theory on an expanding space, and we write down equations for the time
evolution of arbitrary (nonequilibrium) distributions in an expanding
universe. In section 4, we discuss two scenarios whereby quantum
nonequilibrium could exist during inflation: first, nonequilibrium for
large-wavelength modes could survive from a pre-inflationary era, since
under the right conditions relaxation can be suppressed at large wavelengths
on an expanding space; second, nonequilibrium might be generated by novel
gravitational processes at the Planck scale. In section 5, we review the
standard theory of CMB temperature anisotropies, their explanation in terms
of primordial curvature perturbations, and the production of the latter by
inflaton fluctuations during inflation. In section 6, we calculate the time
evolution of quantum nonequilibrium in the Bunch-Davies vacuum on de Sitter
space, and we show that the width $D_{k}(t)$ of the nonequilibrium
distribution for each mode of wave number $k$ remains in a fixed ratio $%
\sqrt{\xi (k)}\equiv D_{k}(t)/\Delta _{k}(t)$ with the equilibrium (quantum)
width $\Delta _{k}(t)$. In section 7, we show how the power spectrum for the
primordial curvature perturbations is corrected by the factor $\xi (k)$.
Some general remarks are made in section 8, concerning the transfer of
microscopic nonequilibrium to cosmological scales, the effective quantum
measurement of the inflaton field during the `quantum-to-classical'
transition, and the weak dependence of our results on the details of
pilot-wave dynamics. In section 9, we use current CMB data to derive an
approximate bound on quantum nonequilibrium during inflation; specifically,
under certain assumptions, we show that the hidden-variable relative entropy 
$S_{\mathrm{hv}}(k)$ (which measures the difference between nonequilibrium
and quantum probabilities for a mode of wave number $k$) satisfies the
approximate bound $\left\vert S_{\mathrm{hv}}(k)\right\vert \lesssim 10^{-2}$
for values of $k$ close to $k_{0}=0.002\ \mathrm{Mpc}^{-1}$. In section 10,
we consider the possibility of a low-power anomaly at large angular scales,
and we discuss how it might arise from a nonequilibrium suppression of
quantum noise ($\xi (k)<1$) in certain regions of $k$-space. In section 11,
we show how nonequilibrium can generate primordial perturbations with
non-random phases and inter-mode correlations. Our conclusions are given in
section 12.

\section{Quantum equilibrium and quantum nonequilibrium}

The notion of quantum nonequilibrium was first discussed in detail in terms
of de Broglie-Bohm theory \cite{AV91a,AV91b,AV92}, and was later generalised
to include all (deterministic) hidden-variables theories \cite%
{AV02a,AV02b,AVSig,PV06}.

Consider, for example, the very simple case of de Broglie-Bohm theory
applied to a single nonrelativistic particle with mass $m$ and no spin. The
wave function{\small \ }$\psi =\left\vert \psi \right\vert e^{iS}$ (with
units $\hbar =1$) acts as a `pilot wave' that determines the velocity of the
particle according to de Broglie's guidance equation $d\mathbf{x}/dt=(1/m)%
\mathbf{\nabla }S$ (or $d\mathbf{x}/dt=(1/m)\func{Im}\left( \mathbf{\nabla }%
\psi /\psi \right) $) --- an equation that determines the trajectory $%
\mathbf{x}(t)$ of the particle, given the initial position $\mathbf{x}(0)$
(assuming that $\psi =\psi (\mathbf{x},t)$ is known for all $\mathbf{x}$ and 
$t$, by solving the Schr\"{o}dinger equation with a given initial wave
function $\psi (\mathbf{x},0)$). Let $\psi $ propagate in free space, then
strike a screen with two slits, and finally strike a backstop where the
particle is detected. The pilot wave undergoes interference upon traversing
the screen. The location $\mathbf{x}(t)$ of the particle at any time $t$ is
determined (in principle) by the initial value $\mathbf{x}(0)$; in
particular, where the particle lands on the backstop is determined by $%
\mathbf{x}(0)$. Because the velocity field $(1/m)\mathbf{\nabla }S$ is equal
to the usual quantum probability current $\mathbf{j}$ divided by the usual
quantum probability density $\left\vert \psi \right\vert ^{2}$, it follows
trivially that an initial ensemble of particles guided by the same pilot
wave $\psi $ and with positions $\mathbf{x}(0)$ distributed according to the
equilibrium rule $\rho (\mathbf{x},0)=\left\vert \psi (\mathbf{x}%
,0)\right\vert ^{2}$ will evolve into an equilibrium distribution $\rho (%
\mathbf{x},t)=\left\vert \psi (\mathbf{x},t)\right\vert ^{2}$ at later
times, resulting in the usual quantum distribution of particles at the
backstop (showing the usual interference pattern). On the other hand, it is
easy to see that in general an initial `nonequilibrium' ensemble with
distribution $\rho (\mathbf{x},0)\neq \left\vert \psi (\mathbf{x}%
,0)\right\vert ^{2}$ results in a non-quantum distribution $\rho (\mathbf{x}%
,t)\neq \left\vert \psi (\mathbf{x},t)\right\vert ^{2}$ at the backstop.
(For example, in the absence of a rapid divergence of neighbouring
trajectories, if $\rho (\mathbf{x},0)$ is concentrated around a single
initial point $\mathbf{x}(0)$ then $\rho (\mathbf{x},t)$ will be
concentrated around a single trajectory $\mathbf{x}(t)$, and the usual
interference pattern will be replaced by a single localised spot.)

The pilot-wave theory of a many-body system was first proposed by de Broglie
at the 1927 Solvay conference \cite{deB28,BV,AV09}. For a system of $n$
(nonrelativistic) particles with positions $\mathbf{x}_{i}(t)$ and masses $%
m_{i}$, de Broglie's law of motion takes the form%
\begin{equation}
\frac{d\mathbf{x}_{i}}{dt}=\frac{1}{m_{i}}\func{Im}\frac{\mathbf{\nabla }%
_{i}\psi }{\psi }=\frac{\mathbf{\nabla }_{i}S}{m_{i}}\ ,  \label{deB}
\end{equation}%
where $\psi =\psi (\mathbf{x}_{1},....,\mathbf{x}_{n},t)$ is the many-body
wave function. De Broglie regarded (\ref{deB}) as expressing a unification
of the principles of Maupertuis and Fermat, resulting in a new form of
dynamics based on velocities \cite{BV}.

Writing the total configuration as $q=(\mathbf{x}_{1},\mathbf{x}_{2},....,%
\mathbf{x}_{n})$, it is again readily shown that for an ensemble of systems
guided by the same wave $\psi $ and with configurations distributed
according to $\rho (q,0)=\left\vert \psi (q,0)\right\vert ^{2}$, the
distribution of configurations at later times will be $\rho (q,t)=\left\vert
\psi (q,t)\right\vert ^{2}$.

As shown in detail by Bohm in 1952 \cite{B52}, the above `de Broglian'
dynamics may be applied to the process of quantum measurement itself, by
treating the system being measured together with the measuring apparatus as
a single many-body system of $n$ particles. The total configuration $q=(%
\mathbf{x}_{1},\mathbf{x}_{2},....,\mathbf{x}_{n})$ then defines the
`pointer position' of the apparatus, as well as defining the configuration
of the measured system. For each run of a quantum experiment, the evolution
is deterministic: the initial conditions $q(0)$, $\psi (q,0)$ determine the
final conditions $q(t)$, $\psi (q,t)$. Over an ensemble of initial
configurations $q(0)$ guided by the same wave function $\psi $, if we assume
the initial quantum equilibrium condition $\rho (q,0)=\left\vert \psi
(q,0)\right\vert ^{2}$, then the statistical distribution of pointer
positions at later times will agree with the predictions of quantum theory.

Schematically, during a standard quantum measurement, the initial packet $%
\psi (q,0)$ on configuration space evolves into a superposition $\psi
(q,t)=\sum_{n}c_{n}\psi _{n}(q,t)$ of terms $\psi _{n}(q,t)$ that separate
with respect to the pointer degrees of freedom (that is, distinct $\psi
_{n}(q,t)$ have negligible overlap with respect to the pointer degrees of
freedom). The final configuration $q(t)$ can then be in (the support of)
only one `branch' of the superposition, say $\psi _{i}(q,t)$. For an initial
equilibrium ensemble, it is readily shown that this occurs with probability $%
\left\vert c_{i}\right\vert ^{2}$, in accordance with the Born rule.
Further, inspection of de Broglie's velocity law (\ref{deB}) shows that the
motion of $q(t)$ will then be affected by $\psi _{i}(q,t)$ alone, resulting
in an effective `reduction' of the wave function.

As in the simple example of a single particle undergoing interference, for a
general quantum measurement the distribution of outcomes depends crucially
on the assumed initial distribution $\rho (q,0)$ of initial configurations $%
q(0)$. For a nonequilibrium ensemble, $\rho (q,0)\neq \left\vert \psi
(q,0)\right\vert ^{2}$, the distribution of quantum measurement outcomes
will generally disagree with the predictions of quantum theory (assuming
that relaxation to equilibrium has not taken place in the meantime --- see
below).

De Broglie's dynamics may be readily applied to fields, where (say for a
scalar field $\phi $) the motion of the field configuration $q(t)=\phi (%
\mathbf{x},t)$ is determined by the Schr\"{o}dinger wave functional $\Psi
\lbrack \phi (\mathbf{x}),t]$. Indeed, for any system with configuration $q$
and Hamiltonian $\hat{H}$, as long as the Schr\"{o}dinger equation $%
i\partial \psi /\partial t=\hat{H}\psi $ for $\psi (q,t)$ has an associated
current $j=j\left[ \psi \right] =j(q,t)$ in configuration space, obeying a
continuity equation%
\begin{equation*}
\frac{\partial \left\vert \psi \right\vert ^{2}}{\partial t}+\nabla \cdot j=0
\end{equation*}%
(with $\nabla \equiv \partial /\partial q$), one may define a de Broglian or
pilot-wave dynamics for the system, by introducing the configuration-space
velocity field%
\begin{equation}
\frac{dq}{dt}=\frac{j}{\left\vert \psi \right\vert ^{2}}\ .  \label{deBgen}
\end{equation}%
Such a velocity field exists, in fact, whenever $\hat{H}$ is given by a
differential operator \cite{SV09}. (In this dynamics, $\psi $ is viewed as a
physical field or `pilot wave' in configuration space, guiding the motion of
an individual system. Note that $\psi $ has no \textit{a priori} connection
with probabilities. Furthermore, because $\psi $ is not an ordinary field in
spacetime, it does not itself carry an energy or momentum density.)

For an ensemble of systems, each with the same wave function $\psi (q,t)$,
we may consider an arbitrary initial distribution $\rho (q,0)\neq \left\vert
\psi (q,0)\right\vert ^{2}$, whose time evolution $\rho (q,t)$ is determined
by the de Broglian velocity field $\dot{q}$ in accordance with the
continuity equation%
\begin{equation*}
\frac{\partial \rho }{\partial t}+\nabla \cdot \left( \rho \dot{q}\right)
=0\ .
\end{equation*}%
Because $\left\vert \psi \right\vert ^{2}$ obeys the same equation, an
initial distribution $\rho (q,0)=\left\vert \psi (q,0)\right\vert ^{2}$
evolves into $\rho (q,t)=\left\vert \psi (q,t)\right\vert ^{2}$. This is the
state of quantum equilibrium, but the theory clearly allows one (in
principle) to consider any initial distribution --- just as classical
mechanics allows one to consider any initial distribution departing from
thermal equilibrium.

It is worth emphasising that pilot-wave theory is a radically new form of
dynamics, very different from classical (Newtonian or Hamiltonian)
mechanics. This was in fact de Broglie's original point of view, but it was
unfortunately obscured by Bohm's pseudo-Newtonian reformulation of the
theory in terms of a law for acceleration (involving a `quantum potential') 
\cite{BV}.

Pilot-wave dynamics is grounded in configuration space, where $\psi $
propagates. While the dynamics is local in configuration space, it is highly
nonlocal when projected down to 3-space (as required by Bell's theorem). For
example, if a particle with position $\mathbf{x}_{1}$ is entangled with a
particle with position $\mathbf{x}_{2}$, then the velocity $\mathbf{\dot{x}}%
_{1}$ depends instantaneously on $\mathbf{x}_{2}$ (no matter how remote $%
\mathbf{x}_{2}$ may be from $\mathbf{x}_{1}$), and changing the local
Hamiltonian at $\mathbf{x}_{2}$ is found to have an instantaneous effect on
the distant velocity $\mathbf{\dot{x}}_{1}$. Such nonlocal effects are
erased upon averaging over an equilibrium ensemble $\rho =\left\vert \psi
\right\vert ^{2}$; but in nonequilibrium, $\rho \neq \left\vert \psi
\right\vert ^{2}$, there are (in general) nonlocal signals at the
statistical level \cite{AV91b,AV92}, suggesting the existence of an
underlying preferred foliation of spacetime \cite{AVSim}.

Pilot-wave dynamics --- as originally formulated by de Broglie --- is also
first order in time in configuration space (rather than in phase space): the
fundamental law of motion determines velocities, not accelerations. This
last feature has important implications for the associated kinematics: for
particles, the natural state of motion is rest (instead of uniform motion in
a straight line), and there is indeed a natural preferred foliation of
spacetime with a fundamental time parameter $t$ (consistent with the
fundamental nonlocality of the theory) \cite{AV97}.

Quantum nonequilibrium may be considered, not only in pilot-wave theory, but
also in any deterministic hidden-variables theory \cite%
{AV02a,AV02b,AVSig,PV06}. For any such theory, given macroscopic
experimental settings $M$, there is a mapping $\omega =\omega (M,\lambda )$
from initial hidden variables $\lambda $ to final outcomes $\omega $ of
quantum measurements. There is also a `quantum equilibrium' probability
measure $\rho _{\mathrm{QT}}(\lambda )$, defined on the set of hidden
variables, that yields quantum probabilities $P_{\mathrm{QT}}(\omega )$ for
the outcomes. (In the case of pilot-wave theory, $\rho _{\mathrm{QT}%
}(\lambda )$ is given by $\rho =\left\vert \psi \right\vert ^{2}$.) Once
such a theory has been constructed, one may consider arbitrary
`nonequilibrium' probability measures $\rho (\lambda )\neq \rho _{\mathrm{QT}%
}(\lambda )$, resulting in outcome probabilities $P(\omega )\neq P_{\mathrm{%
QT}}(\omega )$ that depart from the predictions of quantum theory.

In this paper we shall be studying quantum nonequilibrium in the context of
inflationary cosmology, using the pilot-wave theory of fields as a concrete
example. However, we emphasise that similar studies could be made in any
deterministic hidden-variables theory, simply by making the replacement $%
\rho _{\mathrm{QT}}(\lambda )\rightarrow \rho (\lambda )$.

At present, pilot-wave theory is the only deterministic hidden-variables
theory of broad scope that we possess, though some attempts have been made
to construct alternative theories. For example, in the 1980s, Smolin
attempted to construct a deterministic hidden-variables theory of an $N$%
-body system, based on the classical Hamiltonian dynamics of a certain $%
N\times N$ matrix $M_{ij}(t)$, whose eigenvalues correspond to particle
positions and whose off-diagonal elements correspond to nonlocal hidden
variables associated with pairs of particles \cite{Sm86}. Adopting a
classical action principle for the (deterministic) dynamics of the matrix,
Smolin made a number of assumptions, including a statistical assumption to
the effect that the coarse-grained evolution of the off-diagonal terms
amounts to a Brownian motion. In the limit of a large number $N$ of
particles with masses $m_{i}$, it was shown from these assumptions that the
particle positions also undergo a Brownian motion, that the $i$th particle
current velocity $\mathbf{v}_{i}$ (the average of the mean forward and
backward velocities) is given by a gradient, $\mathbf{v}_{i}=\mathbf{\nabla }%
_{i}S/m_{i}$, where $S$ is a function on configuration space, and that the
complex function $\psi \equiv \sqrt{\rho }e^{iS}$ (where $\rho $ is the
particle probability distribution on configuration space) satisfies the Schr%
\"{o}dinger equation for a many-body nonrelativistic system. Smolin's
strategy was to show that his assumptions led, in the limit of large $N$, to
the basic postulates of Nelson's stochastic mechanics \cite{N66}. As was
already known, in Nelson's theory --- which is based on a form of Brownian
motion subject to special conditions, including the condition that $\mathbf{v%
}_{i}=\mathbf{\nabla }_{i}S/m_{i}$ for some function $S$ --- the derived
quantity $\psi \equiv \sqrt{\rho }e^{iS}$ indeed satisfies the Schr\"{o}%
dinger equation.

More recently, a model similar to the above (though based on the bosonic
part of the classical matrix models used in string and M theory) was again
investigated by Smolin, with similar assumptions and results \cite{Sm02}. In
ref. \cite{Sm86}, it had also been suggested that one might consider a model
in which the off-diagonal matrix elements of $M_{ij}(t)$ are constant, with
fluctuations in a local system arising from the nonlocal transmission of
fluctuations from other particles in remote regions of space. This last
model has recently been recast in terms of the dynamics of a graph with $N$
nodes \cite{MS04}: assuming that the edges of the graph do not evolve in
time, the corresponding adjacency matrix is constant, and is taken to be the
off-diagonal part of matrices $M_{ij}(t)$. Again, as in Smolin's original
model, assumptions are made so as to arrive at Nelson's stochastic mechanics
in some approximation.

However, while it is often claimed that Nelson's theory is empirically
equivalent to quantum theory, unfortunately, as shown by Wallstrom \cite%
{Wall}, the two theories are in fact not equivalent, because Nelson's
function $S$ does not have the specific multivalued structure required for
the phase of a single-valued (and continuous) complex field $\psi $. The Schr%
\"{o}dinger equation is indeed derived, but only for the exceptional set of
wave functions with no nodes, for which the circulation of $\mathbf{\nabla }%
_{i}S$ around all closed curves vanishes. Since almost all wave functions
have nodal points (where $\psi =0$), quantum theory cannot be derived from
Nelson's theory, or from any model that leads to Nelson's theory. (Note that
there is no such problem in pilot-wave theory, where $\psi $ is regarded as
a basic entity.)

Thus, as they stand, the deterministic models of refs. \cite{Sm86,Sm02,MS04}
seem to yield derivations of Nelsonian mechanics, but not of quantum
mechanics. Some basic element is missing. One must somehow ensure that the
circulation of $\mathbf{\nabla }_{i}S$ around nodes of $\rho $ can be
non-zero but always restricted to integer multiples of $2\pi $. (And if one
wishes to derive the wave function, then of course one cannot simply assume
at the outset that $S$ is the phase of a complex-valued field.) Still, if
some way were found to solve Wallstrom's phase problem, then such
derivations of Nelsonian mechanics as an average over a certain statistical
state could again be generalised to arbitrary statistical states, yielding
nonequilibrium departures from quantum theory in the sense considered here.%
\footnote{%
Smolin \cite{Sm06} has attempted to solve Wallstrom's phase problem by
allowing discontinuous wave functions. However (even leaving aside the
resulting divergences for expectation values of quantum observables such as
kinetic energy), Smolin applies his prescription only to the case of a
particle moving on a circle, which is too simple to capture the nature of
the problem raised by Wallstrom. In higher dimensions --- for example even
in two dimensions, and with just one node --- allowing discontinuous wave
functions results in an ill-defined (one-to-many) mapping from Nelsonian
states to quantum states. For a full discussion, see ref. \cite{HVSBK}.}

As another example, Adler \cite{Ad04} has constructed what appears to be a
deterministic hidden-variables theory, in which the parameters $\lambda $
are matrices with Grassmann (even and odd) valued elements, obeying a
generalised form of classical Hamiltonian dynamics. The state of thermal
equilibrium, defined in the usual way on phase space, is argued to lead
(after some approximations) to a quantum-like phenomenology with a dynamical
wave function collapse. The precise nature of Adler's theory seems to
require further elucidation; but if it is indeed a hidden-variables theory
in the sense meant here, then thermal nonequilibrium in Adler's theory
should again correspond to quantum nonequilibrium.

In the author's view, because Hamiltonian dynamics is of second order in
configuration space, it is not a natural framework for nonlocal theories
with a preferred state of rest or preferred slicing of spacetime --- unlike
pilot-wave dynamics, which is first order in configuration space, and which
therefore (as we have mentioned) provides a natural setting for such
theories \cite{AV97}. But even so, the above alternative theories based on
Hamiltonian dynamics do illustrate that the idea of quantum nonequilibrium
is a general one.

For the purposes of this paper, it suffices that there exists at least one
model of quantum nonequilibrium, based on pilot-wave dynamics, that may
serve as a foil against which to test quantum theory. To be able to provide
quantitative bounds on violations of quantum theory in the early universe is
motivation enough to consider models with quantum nonequilibrium. Even so,
before proceeding, let us briefly provide some further motivation for why
quantum nonequilibrium might exist at very early times.

First, it has been shown that in pilot-wave theory the equilibrium state $%
\rho =\left\vert \psi \right\vert ^{2}$ may be understood as arising from a
process of relaxation that is analogous to classical thermal relaxation,
where the former is defined on configuration space rather than on phase
space. The difference between $\rho $ and $\left\vert \psi \right\vert ^{2}$
may be quantified by the $H$-function%
\begin{equation}
H=\int dq\ \rho \ln (\rho /\left\vert \psi \right\vert ^{2})  \label{H}
\end{equation}%
(equal to minus the relative entropy of $\rho $ with respect to $\left\vert
\psi \right\vert ^{2}$), which obeys a coarse-graining $H$-theorem analogous
to the classical one, and where the minimum $H=0$ corresponds to $\rho
=\left\vert \psi \right\vert ^{2}$ \cite{AV91a,AV92,AV01}. Further,
numerical simulations for simple two-dimensional systems \cite{VW05,Sky}
show a remarkably efficient approach to equilibrium, with an approximately
exponential decay of the coarse-grained $H$-function, $\bar{H}(t)\rightarrow
0$, and a corresponding coarse-grained relaxation $\bar{\rho}\rightarrow 
\overline{\left\vert \psi \right\vert ^{2}}$ (assuming appropriate initial
conditions for $\rho $ and $\psi $).\footnote{%
The understanding of relaxation in pilot-wave theory is subject to the usual
caveats --- familiar from classical statistical mechanics --- associated
with initial conditions and time reversal. For detailed discussions of this
point, see refs. \cite{AV92,AV96,AV01,VW05}.} Because all the systems we
have access to (such as hydrogen atoms in the laboratory) have a long and
violent astrophysical history, we would then \textit{expect} to see quantum
equilibrium in these systems. While it is logically possible, of course,
that the universe was simply born in a state of quantum equilibrium, it
seems more natural to consider that the equilibrium we see today arose from
relaxation processes in the remote past \cite{AV96,AV01}, in which case the
very early universe is the natural place to look for nonequilibrium
phenomena.

Second, an appealing feature of this picture concerns the status of locality
in physics. It may be shown that quantum nonequilibrium for entangled
systems leads to nonlocal signals at the statistical level, in pilot-wave
theory (as already mentioned) and indeed in any deterministic
hidden-variables theory; while in equilibrium, the underlying nonlocal
effects cancel out at the statistical level \cite%
{AV91b,AV92,AV02a,AV02b,AV09}. Locality is therefore a contingency (or
emergent feature) of the equilibrium state. Similarly, standard
uncertainty-principle limitations on measurements are also contingencies of
equilibrium \cite{AV91b,AV92,AVPr02,PV06}. These results provide an
explanation for the otherwise mysterious `conspiracy' in the foundations of
current physics, according to which (roughly speaking) quantum noise and the
uncertainty principle prevent us from using quantum nonlocality for
practical nonlocal signalling. From the above perspective, this `conspiracy'
is not part of the laws of physics, but merely a contingent feature of the
equilibrium state (much as the inability to convert heat into work, in a
state of global thermal equilibrium, is not a law of physics but a
contingency of the state). On this view, quantum physics is merely the
effective description of a particular state --- just as, for example, the
standard model of particle physics is merely the effective description of
(perturbations around) a particular vacuum state (arising from spontaneous
symmetry breaking). If one takes this view seriously, it suggests that
nonequilibrium phenomena should exist somewhere (or some time) in our
universe. And again, the early universe seems the natural place to look.

Quantum nonequilibrium at very early times may also be motivated by the
cosmological horizon problem, which may be avoided by the explicit
nonlocality associated with nonequilibrium \cite{AV91b,AV92,AV96,AV02b} ---
see section 4.1.

Finally, if one takes de Broglie-Bohm theory seriously, one should take the
possibility of nonequilibrium seriously as well, since it is only in
nonequilibrium that the underlying details of the theory become visible (via
measurements more accurate than those allowed by quantum theory \cite%
{AVPr02,PV06}). If instead the universe is always and everywhere in quantum
equilibrium, the details of de Broglie-Bohm trajectories will be forever
shielded from experimental tests, and de Broglie-Bohm theory itself would be
unacceptable as a scientific theory.

For the above reasons, then, we are led to consider the hypothesis of
quantum nonequilibrium at or close to the big bang \cite%
{AV91a,AV91b,AV92,AV96,AV01,AV07,AV09}. It is the purpose of this paper to
show that inflationary cosmology provides a means of testing this
hypothesis, through precision measurements of the cosmic microwave
background.

\section{Pilot-wave field theory on expanding space}

For simplicity we restrict ourselves to a flat metric,%
\begin{equation}
d\tau ^{2}=dt^{2}-a^{2}d\mathbf{x}^{2}\ ,  \label{metric}
\end{equation}%
where again $a(t)$ is the scale factor, with Hubble parameter $H\equiv \dot{a%
}/a$. As is customary, we take $a_{0}=1$ today (at time $t_{0}$), so that $|d%
\mathbf{x}|$ is a comoving distance (or proper distance today).

A free (minimally-coupled) massless scalar field $\phi $ has a Lagrangian
density $\mathcal{L}=\frac{1}{2}g^{1/2}\partial _{\alpha }\phi \partial
^{\alpha }\phi $ or%
\begin{equation}
\mathcal{L}=\tfrac{1}{2}a^{3}\dot{\phi}^{2}-\tfrac{1}{2}a(\mathbf{\nabla }%
\phi )^{2}\ ,  \label{Lagden}
\end{equation}%
with an action $\int dt\int d^{3}\mathbf{x}\;\mathcal{L}$ (where $\mathbf{x}$
are comoving coordinates). This implies a canonical momentum density $\pi
=\partial \mathcal{L}/\partial \dot{\phi}=a^{3}\dot{\phi}$ and a Hamiltonian
density%
\begin{equation}
\mathcal{H}=\tfrac{1}{2}\frac{\pi ^{2}}{a^{3}}+\tfrac{1}{2}a(\mathbf{\nabla }%
\phi )^{2}\ .  \label{Hamden}
\end{equation}%
The equations of motion $\dot{\phi}=\delta H/\delta \pi $, $\dot{\pi}%
=-\delta H/\delta \phi $ (with $H=\int d^{3}\mathbf{x}\;\mathcal{H}$) lead
to the classical wave equation%
\begin{equation}
\ddot{\phi}+\frac{3\dot{a}}{a}\dot{\phi}-\frac{1}{a^{2}}\nabla ^{2}\phi =0\ .
\label{weqn}
\end{equation}

Pilot-wave field theory is defined in terms of the functional Schr\"{o}%
dinger picture, with a preferred foliation of spacetime \cite%
{AV92,AV96,B52,BH84,BHK87,Holl93,BandH,Kal94,StruyPR}. For an expanding
universe with metric (\ref{metric}), containing a scalar field $\phi $ with
Hamiltonian density (\ref{Hamden}), a general wave functional $\Psi \lbrack
\phi ,t]=\langle \phi (\mathbf{x})|\Psi (t)\rangle $ (where $|\phi (\mathbf{x%
})\rangle $ is a field eigenstate) satisfies the functional Schr\"{o}dinger
equation\footnote{%
As usual in this context, some sort of regularisation is implicitly assumed.}%
\begin{equation}
i\frac{\partial \Psi }{\partial t}=\int d^{3}\mathbf{x}\;\left( -\frac{1}{%
2a^{3}}\frac{\delta ^{2}}{\delta \phi ^{2}}+\frac{1}{2}a(\mathbf{\nabla }%
\phi )^{2}\right) \Psi  \label{Sch1}
\end{equation}%
(with the usual realisations $\hat{\phi}\rightarrow \phi $, $\hat{\pi}%
\rightarrow -i\delta /\delta \phi $). This implies the continuity equation%
\begin{equation}
\frac{\partial \left\vert \Psi \right\vert ^{2}}{\partial t}+\int d^{3}x\;%
\frac{\delta }{\delta \phi }\left( \left\vert \Psi \right\vert ^{2}\frac{1}{%
a^{3}}\frac{\delta S}{\delta \phi }\right) =0  \label{Cont1}
\end{equation}%
(where $\Psi =\left\vert \Psi \right\vert e^{iS}$), from which one may
identify the de Broglie velocity%
\begin{equation}
\frac{\partial \phi }{\partial t}=\frac{1}{a^{3}}\frac{\delta S}{\delta \phi 
}  \label{deB1}
\end{equation}%
for an individual field configuration. Here, again, $\Psi $ is interpreted
as a physical field in configuration space, guiding the evolution of an
individual field $\phi (\mathbf{x},t)$ in 3-space. (Note that $S$ is defined
only locally, as $S=\func{Im}\ln \Psi $. One may equally write (\ref{deB1})
as $\frac{\partial \phi }{\partial t}=\frac{1}{a^{3}}\func{Im}\frac{1}{\Psi }%
\frac{\delta \Psi }{\delta \phi }$, without mentioning $S$.)

A similar construction may be given in any globally-hyperbolic spacetime, by
choosing a preferred foliation \cite{AVBHs}. Thus there is no need for
spatial homogeneity.

Over an ensemble of field configurations guided by the same pilot wave $\Psi 
$, there will be some (in principle, arbitrary) initial distribution $P[\phi
,t_{i}]$, whose time evolution $P[\phi ,t]$ will be determined by%
\begin{equation}
\frac{\partial P}{\partial t}+\int d^{3}x\;\frac{\delta }{\delta \phi }%
\left( P\frac{1}{a^{3}}\frac{\delta S}{\delta \phi }\right) =0\ .
\label{ContP1}
\end{equation}%
If $P[\phi ,t_{i}]=\left\vert \Psi \lbrack \phi ,t_{i}]\right\vert ^{2}$,
then $P[\phi ,t]=\left\vert \Psi \lbrack \phi ,t]\right\vert ^{2}$ for all $%
t $, and empirical agreement is obtained with standard quantum field theory 
\cite{B52,BHK87,Holl93,BandH,Kal94,StruyPR}. On the other hand, for an
initial nonequilibrium distribution $P[\phi ,t_{i}]\neq \left\vert \Psi
\lbrack \phi ,t_{i}]\right\vert ^{2}$, for as long as $P$ remains in
nonequilibrium, the predicted statistics will generally differ from those of
quantum field theory. In any case, whatever form $P$ may take (equilibrium
or nonequilibrium), its time evolution will be given by (\ref{ContP1}).

It will prove convenient to rewrite the dynamics in Fourier space.
Expressing $\phi (\mathbf{x})$ in terms of its Fourier components%
\begin{equation*}
\phi _{\mathbf{k}}=\frac{1}{(2\pi )^{3/2}}\int d^{3}\mathbf{x}\;\phi (%
\mathbf{x})e^{-i\mathbf{k}\cdot \mathbf{x}}\ ,
\end{equation*}%
and writing%
\begin{equation*}
\phi _{\mathbf{k}}=\frac{\sqrt{V}}{(2\pi )^{3/2}}\left( q_{\mathbf{k}1}+iq_{%
\mathbf{k}2}\right)
\end{equation*}%
for real $q_{\mathbf{k}r}$ ($r=1$, $2$), where $V$ is a box normalisation
volume, the Lagrangian $L=\int d^{3}\mathbf{x}\;\mathcal{L}$ becomes%
\begin{equation*}
L=\sum_{\mathbf{k}r}\frac{1}{2}\left( a^{3}\dot{q}_{\mathbf{k}%
r}^{2}-ak^{2}q_{\mathbf{k}r}^{2}\right) \ .
\end{equation*}%
(For $V\rightarrow \infty $, $\frac{1}{V}\sum_{\mathbf{k}}\rightarrow \frac{1%
}{(2\pi )^{3}}\int d^{3}\mathbf{k}$ and $V\delta _{\mathbf{kk%
{\acute{}}%
}}\rightarrow (2\pi )^{3}\delta ^{3}(\mathbf{k}-\mathbf{k}%
{\acute{}}%
)$. The reality of $\phi $ requires $\phi _{\mathbf{k}}^{\ast }=\phi _{-%
\mathbf{k}}$ or $q_{\mathbf{k}1}=q_{-\mathbf{k}1}$, $q_{\mathbf{k}2}=-q_{-%
\mathbf{k}2}$, so that a sum over physical degrees of freedom should be
restricted to half the values of $\mathbf{k}$.) Introducing the canonical
momenta%
\begin{equation*}
\pi _{\mathbf{k}r}\equiv \frac{\partial L}{\partial \dot{q}_{\mathbf{k}r}}%
=a^{3}\dot{q}_{\mathbf{k}r}\ ,
\end{equation*}%
the Hamiltonian becomes%
\begin{equation*}
H=\sum_{\mathbf{k}r}\left( \frac{1}{2a^{3}}\pi _{\mathbf{k}r}^{2}+\frac{1}{2}%
ak^{2}q_{\mathbf{k}r}^{2}\right) \ .
\end{equation*}

The Schr\"{o}dinger equation for $\Psi =\Psi \lbrack q_{\mathbf{k}r},t]$ is
then%
\begin{equation}
i\frac{\partial \Psi }{\partial t}=\sum_{\mathbf{k}r}\left( -\frac{1}{2a^{3}}%
\frac{\partial ^{2}}{\partial q_{\mathbf{k}r}^{2}}+\frac{1}{2}ak^{2}q_{%
\mathbf{k}r}^{2}\right) \Psi \ ,  \label{Sch2}
\end{equation}%
which implies the continuity equation%
\begin{equation}
\frac{\partial \left\vert \Psi \right\vert ^{2}}{\partial t}+\sum_{\mathbf{k}%
r}\frac{\partial }{\partial q_{\mathbf{k}r}}\left( \left\vert \Psi
\right\vert ^{2}\frac{1}{a^{3}}\frac{\partial S}{\partial q_{\mathbf{k}r}}%
\right) =0  \label{Cont2}
\end{equation}%
and the de Broglie velocities%
\begin{equation}
\frac{dq_{\mathbf{k}r}}{dt}=\frac{1}{a^{3}}\frac{\partial S}{\partial q_{%
\mathbf{k}r}}  \label{deB2}
\end{equation}%
(again with $\Psi =\left\vert \Psi \right\vert e^{iS}$). The time evolution
of an arbitrary distribution $P[q_{\mathbf{k}r},t]$ will then be given by%
\begin{equation}
\frac{\partial P}{\partial t}+\sum_{\mathbf{k}r}\frac{\partial }{\partial q_{%
\mathbf{k}r}}\left( P\frac{1}{a^{3}}\frac{\partial S}{\partial q_{\mathbf{k}%
r}}\right) =0\ .  \label{ContP2}
\end{equation}

For product states%
\begin{equation}
\Psi \lbrack q_{\mathbf{k}r},t]=\prod\limits_{\mathbf{k}r}\psi _{\mathbf{k}%
r}(q_{\mathbf{k}r},t)  \label{Psiprod}
\end{equation}%
(such as the Bunch-Davies vacuum during inflation), the wave function $\psi
_{\mathbf{k}r}$ for a single mode $\mathbf{k}r$ satisfies%
\begin{equation}
i\frac{\partial \psi _{\mathbf{k}r}}{\partial t}=\left( -\frac{1}{2a^{3}}%
\frac{\partial ^{2}}{\partial q_{\mathbf{k}r}^{2}}+\frac{1}{2}ak^{2}q_{%
\mathbf{k}r}^{2}\right) \psi _{\mathbf{k}r}\ .  \label{Sch3}
\end{equation}%
Writing $\psi _{\mathbf{k}r}=\left\vert \psi _{\mathbf{k}r}\right\vert
e^{is_{\mathbf{k}r}}$ (where $S=\sum_{\mathbf{k}r}s_{\mathbf{k}r}$), the de
Broglie velocity for $q_{\mathbf{k}r}$ is then%
\begin{equation}
\frac{dq_{\mathbf{k}r}}{dt}=\frac{1}{a^{3}}\frac{\partial s_{\mathbf{k}r}}{%
\partial q_{\mathbf{k}r}}\ .  \label{deB3}
\end{equation}%
If the initial distribution $P[q_{\mathbf{k}r},t_{i}]$ also takes the
product form%
\begin{equation}
P[q_{\mathbf{k}r},t_{i}]=\prod\limits_{\mathbf{k}r}\rho _{\mathbf{k}r}(q_{%
\mathbf{k}r},t_{i})\ ,  \label{Pprod}
\end{equation}%
then the time evolution of $\rho _{\mathbf{k}r}(q_{\mathbf{k}r},t)$ will be
given by%
\begin{equation}
\frac{\partial \rho _{\mathbf{k}r}}{\partial t}+\frac{\partial }{\partial q_{%
\mathbf{k}r}}\left( \rho _{\mathbf{k}r}\frac{1}{a^{3}}\frac{\partial s_{%
\mathbf{k}r}}{\partial q_{\mathbf{k}r}}\right) =0\ .  \label{Cont3}
\end{equation}

Note that the factorisability condition (\ref{Pprod}) for the probability
distribution $P$ is logically independent of the factorisability condition (%
\ref{Psiprod}) for the pilot wave $\Psi $. Thus, even for a vacuum state, in
nonequilibrium it is still possible to have inter-mode correlations. For
simplicity, in section 6, we shall restrict ourselves to the case of
uncorrelated nonequilibrium modes. The correlated case is discussed in
section 11.

\section{Quantum nonequilibrium in the very early universe}

In this paper, the focus is on setting experimental bounds on possible
violations of quantum theory during inflation. Before proceeding with this,
however, let us indicate how one might (in future work) be able to predict
details of such violations. The scenarios sketched in this section also
serve to give a preliminary idea of the kinds of violations one might expect
to find.

\subsection{Relic nonequilibrium from a pre-inflationary era}

One reason to expect early nonequilibrium to exist is that, as sketched in
section 2, according to de Broglie-Bohm theory ordinary matter corresponds
to a `quantum equilibrium phase', and it is natural to suppose that this
equilibrium state emerged from the violence of the big bang.

Another reason is that nonequilibrium at very early times would unleash the
nonlocality inherent in all hidden-variables theories, thereby evading the
horizon problem associated with an early Friedmann expansion (if there was
one). For $a\propto t^{1/2}$ the horizon distance is (with $c=1$)%
\begin{equation*}
l_{\mathrm{h}}(t)=a(t)\int_{0}^{t}\frac{dt%
{\acute{}}%
}{a(t%
{\acute{}}%
)}=2t\ ,
\end{equation*}%
and for any two comoving points separated by a coordinate distance $%
\left\vert \Delta \mathbf{x}\right\vert $, we have $l_{\mathrm{h}%
}(t)<<a(t)\left\vert \Delta \mathbf{x}\right\vert $ for sufficiently small $%
t $. On this basis it has been widely argued that early homogeneity --- over
seemingly causally-disconnected domains --- is unnatural and puzzling.%
\footnote{%
Note, however, that the existence of the puzzle depends on assuming a
classical Friedmann expansion $a\propto t^{1/2}$ all the way back to $t=0$.}
As we have mentioned, the hypothesis of quantum nonequilibrium at the big
bang was originally introduced partly to solve this problem \cite%
{AV91b,AV92,AV96,AV02b}. For the above scalar field, for example, a generic
wave functional $\Psi $ will be entangled across space, so that the field
velocity%
\begin{equation*}
\frac{\partial \phi (\mathbf{x},t)}{\partial t}=\frac{1}{a^{3}}\frac{\delta
S[\phi ,t]}{\delta \phi (\mathbf{x})}
\end{equation*}%
at a point $\mathbf{x}$ will depend on instantaneous values of the field at
remote points $\mathbf{x}%
{\acute{}}%
\neq \mathbf{x}$, and in nonequilibrium this nonlocal dependence will not be
hidden by statistical noise (as it is in quantum theory). Of course the
horizon problem was also one of the historical motivations for introducing
inflation: the period of exponential expansion ensures that our observable
region originates from within a single causal patch \cite{G81}. However,
even in an inflationary context, it appears that some models require
homogeneity as an initial condition in order for inflation to begin \cite%
{VT00}. Therefore, it is possible that consideration of a pre-inflationary
era will revive the horizon problem, and that some form of early nonlocality
may provide a resolution. The nonlocality could be generated by quantum
nonequilibrium, or perhaps by some other means (other proposals include
topological fluctuations \cite{H82} and an increased speed of light at early
times \cite{VSL}).

If we then assume --- for whatever reason --- that the pre-inflationary
universe was in a state of quantum nonequilibrium, the question is how the
nonequilibrium will evolve in time, and in particular, whether any of it
will survive until entry into the inflationary era. To address this
question, let us first summarise what is known so far about relaxation in
pilot-wave theory.

As already mentioned, numerical simulations for simple two-dimensional
systems show an efficient relaxation, with an approximately exponential
decay of the coarse-grained $H$-function $\bar{H}(t)$ \cite{VW05,Sky}.
Specifically, for an ensemble of nonrelativistic particles in a
two-dimensional box (on a static spacetime background), with a wave function
consisting of a superposition of the first 16 modes, it was found that $\bar{%
H}(t)\approx \bar{H}_{0}e^{-t/t_{\mathrm{c}}}$ where, as discussed in ref. 
\cite{VW05}, the timescale $t_{\mathrm{c}}$ coincides approximately with a
theoretical relaxation timescale $\tau $ defined by $1/\tau ^{2}\equiv -(1/%
\bar{H})d^{2}\bar{H}/dt^{2}$ \cite{AV92}, which under certain conditions may
be roughly estimated as \cite{AV01,VW05,AVRelSupp}%
\begin{equation*}
\tau \sim \Delta t\equiv 1/\Delta E\ ,
\end{equation*}%
where $\Delta E$ is the quantum energy spread and $\Delta t$ is the usual
quantum timescale over which the wave function evolves.

Similar results have been obtained for nonrelativistic particles in a
two-dimensional harmonic oscillator potential \cite{Sky}, a case that has
immediate implications for the field theory of a single decoupled mode $%
\mathbf{k}$.

Writing $\Psi =\psi _{\mathbf{k}}(q_{\mathbf{k}1},q_{\mathbf{k}%
2},t)\varkappa $, where $\varkappa $ depends only on degrees of freedom for
modes $\mathbf{k}%
{\acute{}}%
\neq \mathbf{k}$, equations (\ref{Sch2}) and (\ref{deB2}) imply that the
wave function $\psi _{\mathbf{k}}$ satisfies%
\begin{equation}
i\frac{\partial \psi _{\mathbf{k}}}{\partial t}=-\frac{1}{2a^{3}}\left( 
\frac{\partial ^{2}}{\partial q_{\mathbf{k}1}^{2}}+\frac{\partial ^{2}}{%
\partial q_{\mathbf{k}2}^{2}}\right) \psi _{\mathbf{k}}+\frac{1}{2}%
ak^{2}\left( q_{\mathbf{k}1}^{2}+q_{\mathbf{k}2}^{2}\right) \psi _{\mathbf{k}%
}\ ,  \label{AA}
\end{equation}%
while the de Broglie velocities for $(q_{\mathbf{k}1},q_{\mathbf{k}2})$ are%
\begin{equation}
\dot{q}_{\mathbf{k}1}=\frac{1}{a^{3}}\frac{\partial s_{\mathbf{k}}}{\partial
q_{\mathbf{k}1}},\ \ \ \ \dot{q}_{\mathbf{k}2}=\frac{1}{a^{3}}\frac{\partial
s_{\mathbf{k}}}{\partial q_{\mathbf{k}2}}  \label{BB}
\end{equation}%
(with $\psi _{\mathbf{k}}=\left\vert \psi _{\mathbf{k}}\right\vert e^{is_{%
\mathbf{k}}}$). The marginal distribution $\rho _{\mathbf{k}}(q_{\mathbf{k}%
1},q_{\mathbf{k}2},t)$ will evolve according to%
\begin{equation}
\frac{\partial \rho _{\mathbf{k}}}{\partial t}+\sum_{r=1,\ 2}\frac{\partial 
}{\partial q_{\mathbf{k}r}}\left( \rho _{\mathbf{k}}\frac{1}{a^{3}}\frac{%
\partial s_{\mathbf{k}}}{\partial q_{\mathbf{k}r}}\right) =0\ .  \label{CC}
\end{equation}%
As discussed elsewhere \cite{AV07,AVRelSupp}, these are identical to the
equations of pilot-wave dynamics for an ensemble of nonrelativistic
particles with time-dependent `mass' $m=a^{3}$, moving in the
two-dimensional $q_{\mathbf{k}1}-q_{\mathbf{k}2}$ plane, in a harmonic
oscillator potential of time-dependent angular frequency $\omega =k/a$. In
the short-wavelength limit, $\lambda _{\mathrm{phys}}<<\Delta n_{\mathbf{k}%
}\cdot H^{-1}$ (where $n_{\mathbf{k}}=n_{\mathbf{k}1}+n_{\mathbf{k}2}$ is
the sum of the occupation numbers for modes $\mathbf{k}1$ and $\mathbf{k}2$%
), and over timescales $\Delta t\equiv 1/\Delta E_{\mathbf{k}}<<H^{-1}$ (for
which $a$ is approximately constant), the above equations reduce to those
for a decoupled mode $\mathbf{k}$ on Minkowski spacetime \cite{AVRelSupp}.
These limiting equations are in turn just those of pilot-wave dynamics for
an ensemble of nonrelativistic particles of constant mass $m=a^{3}$ in a
two-dimensional harmonic oscillator potential of constant angular frequency $%
\omega =k/a$. The numerical results for this last case \cite{Sky} show that,
in the Minkowski limit, for a decoupled mode $\mathbf{k}$ in a superposition
of many different states of definite occupation number, one will obtain
relaxation $\rho _{\mathbf{k}}(q_{\mathbf{k}1},q_{\mathbf{k}2},t)\rightarrow
\left\vert \psi _{\mathbf{k}}(q_{\mathbf{k}1},q_{\mathbf{k}2},t)\right\vert
^{2}$ (on a coarse-grained level, assuming appropriate initial conditions),
on a timescale $\tau _{\mathbf{k}}$ of order%
\begin{equation*}
\tau _{\mathbf{k}}\sim \frac{1}{\Delta E_{\mathbf{k}}}\ .
\end{equation*}

If, in the Minkowski limit, relaxation occurs so efficiently for a single
decoupled mode, then we may reasonably expect that for a realistic entangled
quantum state --- in some pre-inflationary era --- relaxation will occur at
least as efficiently. One might then conclude that, even if there is initial
nonequilibrium, it will have relaxed away by the time inflation begins.
However, before drawing definite conclusions, one must first consider the
possible effect of spatial expansion on the relaxation process. One finds,
in particular, that the character of the evolution can be very different in
the long-wavelength limit.

In the case of a decoupled mode on expanding space, described by equations (%
\ref{AA})--(\ref{CC}), it is found \cite{AVRelSupp} that in the
long-wavelength limit, $\lambda _{\mathrm{phys}}>>\Delta n_{\mathbf{k}}\cdot
H^{-1}$, the wave function $\psi _{\mathbf{k}}$ is approximately static ---
or `frozen' --- over timescales $\sim H^{-1}$. Furthermore, one expects that
the trajectories $(q_{\mathbf{k}1}(t),q_{\mathbf{k}2}(t))$ will be frozen
over timescales $\sim H^{-1}$, in which case an arbitrary nonequilibrium
distribution $\rho _{\mathbf{k}}\neq \left\vert \psi _{\mathbf{k}%
}\right\vert ^{2}$ will also be frozen over timescales $\sim H^{-1}$. (This
is of course reminiscent of the freezing of super-Hubble modes in the theory
of cosmological perturbations \cite{LL00,Pad93}.) It then begins to appear
possible that the normal process of relaxation to quantum equilibrium could
be suppressed for long-wavelength modes in a pre-inflationary era, and that
remnants of initial nonequilibrium could survive up to the beginning of
inflation.

That this is indeed possible has been shown \cite{AVRelSupp} by deriving a
general and rigorous condition for the freezing of quantum nonequilibrium, a
condition applicable to an arbitrary time interval $[t_{i},t_{f}]$ and to
any (generally entangled) quantum state of a scalar field. (The condition
may also be applied to mixed states and to interacting fields.) The
condition is obtained by considering the displacements of the de
Broglie-Bohm trajectories over the time interval $[t_{i},t_{f}]$. It is
found that, for a pure subensemble with (time-dependent) mean occupation
numbers $\left\langle \hat{n}_{\mathbf{k}r}\right\rangle $, nonequilibrium
will be frozen (or at least partially frozen) for modes with wave number $k$
if the time evolution of $\left\langle \hat{n}_{\mathbf{k}r}\right\rangle $
satisfies the `freezing inequality' \cite{AVRelSupp}%
\begin{equation}
\frac{1}{k}>4a_{f}\sqrt{\left\langle \hat{n}_{\mathbf{k}r}\right\rangle
_{f}+1/2}\int_{t_{i}}^{t_{f}}dt\ \frac{1}{a^{2}}\sqrt{\left\langle \hat{n}_{%
\mathbf{k}r}\right\rangle +1/2}\ .  \label{fr}
\end{equation}%
For a radiation-dominated expansion on $[t_{i},t_{f}]$, with $%
a(t)=a_{f}(t/t_{f})^{1/2}$, this inequality implies that (using $%
\left\langle \hat{n}_{\mathbf{k}r}\right\rangle \geq 0$)%
\begin{equation}
\lambda _{\mathrm{phys}}(t_{f})>2\pi H_{f}^{-1}\ln (t_{f}/t_{i})\ ,
\label{lb}
\end{equation}%
where $H_{f}^{-1}=2t_{f}$ and the right-hand side is larger than $H_{f}^{-1}$
if $t_{f}\gtrsim (1.17)t_{i}$. Thus, in a radiation-dominated expansion, if
the freezing inequality (\ref{fr}) is satisfied, the corresponding modes
must be super-Hubble \cite{AVRelSupp}.

We are now in a position to begin to address the question of whether or not
very early nonequilibrium in a pre-inflationary era could survive until the
onset of inflation itself. The above considerations show that, for
short-wavelength modes, any initial quantum nonequilibrium is likely to be
rapidly destroyed during a pre-inflationary phase. On the other hand,
relaxation can be suppressed for long-wavelength modes --- if the freezing
inequality (\ref{fr})\ is satisfied --- and it is then possible that these
modes (for whatever fields may be present) will still be in nonequilibrium
at the onset of inflation.

Denoting, for a moment, the (approximately) constant Hubble radius during
inflation by $H_{\mathrm{\inf }}^{-1}$, relevant cosmological fluctuations
originate from inside $H_{\mathrm{\inf }}^{-1}$. For some of these modes to
be out of equilibrium, they must have evolved from modes that were outside
the Hubble radius in the (presumably radiation-dominated) pre-inflationary
phase. Therefore, for this scenario to work, some pre-inflationary
nonequilibrium modes must enter the Hubble radius during the transition to
the inflationary phase, and they must avoid complete relaxation to
equilibrium by the time inflation begins. (As we shall see in section 6,
relaxation does not occur during inflation itself.) Now, modes of physical
wavelength $\lambda _{\mathrm{phys}}=a\lambda $ can enter the Hubble radius $%
H^{-1}=a/\dot{a}$ only if $\lambda _{\mathrm{phys}}$ increases more slowly
than does $H^{-1}$, that is, only if the comoving Hubble radius $H^{-1}/a=1/%
\dot{a}$ increases --- as occurs for a decelerating universe, $\ddot{a}<0$
(which, from the Friedmann equation $\ddot{a}/a=-(4\pi G/3)(\rho +3p)$,
requires that the energy density $\rho $ and pressure $p$ satisfy $\rho
+3p>0 $).

During a decelerating pre-inflationary phase, then, any frozen
nonequilibrium modes at super-Hubble radii can enter the Hubble radius. Once
they do so, they are likely to begin to relax to equilibrium. For all modes
that are inside $H_{\mathrm{\inf }}^{-1}$ at the onset of inflation, some
time will necessarily have been spent in what might be crudely termed the
`relaxation zone', with $\lambda _{\mathrm{phys}}\lesssim H^{-1}$, during
the pre-inflationary phase. For example, for a radiation-dominated
pre-inflationary phase (starting at some initial time $t_{i}$) that makes an
abrupt transition to an inflationary phase at $t=t_{f}$, we have $%
a=a_{f}(t/t_{f})^{1/2}$ and $H^{-1}=2t$ (on $[t_{i},t_{f}]$), and a mode of
comoving wavelength $\lambda $ enters the Hubble radius ($a\lambda \sim
H^{-1}$) at a time $t_{\mathrm{enter}}(\lambda )\sim a_{f}^{2}\lambda
^{2}/t_{f}$, so that the time spent in the relaxation zone is%
\begin{equation*}
\Delta t_{\mathrm{relax}}(\lambda )=(t_{f}-t_{\mathrm{enter}})\sim
t_{f}(1-a_{f}^{2}\lambda ^{2}/t_{f}^{2})\ .
\end{equation*}%
There can be significant residual nonequilibrium at the beginning of
inflation, provided the `no relaxation' condition%
\begin{equation}
\Delta t_{\mathrm{relax}}(\lambda )\lesssim \tau (\lambda )  \label{norelax}
\end{equation}%
is satisfied, where $\tau (\lambda )$ is again a relaxation timescale as
defined above (and where $\tau (\lambda )$ may be evaluated at the
intermediate time $t_{\mathrm{enter}}+\frac{1}{2}(t_{f}-t_{\mathrm{enter}})$%
). Because $\tau (\lambda )$ will depend on the wave functional, a proper
calculation of $\tau (\lambda )$ requires a specific model of the
pre-inflationary phase.

Given a specific form for the function $\tau (\lambda )$, the condition (\ref%
{norelax}) will determine a range of wavelengths $\lambda $ for which
residual nonequilibrium may reasonably be expected to have survived from the
pre-inflationary era. Because pre-inflationary modes with larger values of $%
\lambda $ enter the Hubble radius later and so spend less time in the
relaxation zone, the condition (\ref{norelax}) will presumably imply that
residual nonequilibrium will be possible for $\lambda $ larger than some
infra-red cutoff $\lambda _{\mathrm{c}}$. (The scenario might be improved
if, during the transition from a pre-inflationary to an inflationary phase,
the Hubble radius was a rapidly increasing function of the scale factor ($%
dH^{-1}/da>>H^{-1}/a$). For then super-Hubble nonequilibrium modes could be
pushed far inside the Hubble radius in a short time.)

We hope that future work, based on a specific pre-inflationary model, will
yield a prediction for the infra-red cutoff $\lambda _{\mathrm{c}}$. It is
of course possible that $\lambda _{\mathrm{c}}$ will turn out to be so much
larger than today's Hubble radius that it yields a negligible effect on CMB
predictions (as could occur if the relaxation timescale $\tau (\lambda )$ is
too short, or if the number of inflationary e-folds is too large). This
remains to be seen.

In this paper, the focus is on `phenomenology': we simply assume that some
modes could be in quantum nonequilibrium at the beginning of the
inflationary phase, and we show how CMB data may be used to set experimental
bounds on such nonequilibrium. Still, the above preliminary reasoning
already suggests that if there is residual nonequilibrium from a
pre-inflationary phase, then we should expect to find it at large
wavelengths, beyond some cutoff $\lambda _{\mathrm{c}}$.

\subsection{Possible production of nonequilibrium at the Planck scale}

We have discussed whether nonequilibrium might have survived into the
inflationary phase, on the assumption that there was nonequilibrium in some
pre-inflationary era. Another question is whether nonequilibrium might be 
\textit{generated} during (or indeed even before) the inflationary era.

The creation of quantum nonequilibrium from a prior equilibrium state is
impossible in standard de Broglie-Bohm theory (leaving aside extremely rare
fluctuations \cite{AV92}), though it might occur in alternative
hidden-variables models --- for example, in models that deviate from quantum
theory for processes taking place over very short timescales \cite{BB66}.
But even in de Broglie-Bohm theory, it does not seem entirely clear if we
know how to incorporate gravitation \cite{HVSBK} (see, however, ref. \cite%
{PNSS02}). It is therefore conceivable that effects involving gravity are
able to upset the equilibrium state. In particular, as has been discussed at
length elsewhere, it is not unreasonable to propose that quantum
nonequilibrium can be generated by the formation and evaporation of a black
hole \cite{AVBHs,AV07}.

This proposal is motivated by the (controversial) question of information
loss in black holes. In the standard picture of black-hole formation and
evaporation, it appears that a closed system can evolve from an initial pure
state to a final mixed state, thereby violating ordinary quantum theory \cite%
{H76}. Further, because the final state describes thermal radiation that
depends on the initial mass of the hole but not on the details of the
initial state, it is impossible even in principle to retrodict the initial
state from the final state. While Hawking's original argument for
information loss remains controversial, a new approach to avoiding
information loss invokes the possible existence of quantum nonequilibrium in
the outgoing radiation, which could then carry more information than
ordinary radiation can in a conventional (mixed) quantum state. A mechanism
for the creation of such nonequilibrium has been outlined \cite{AVBHs},
involving an assumed nonequilibrium behind the horizon (presumably near the
singularity) that is transferred to the exterior region by the entanglement
between the ingoing and outgoing modes of the Hawking radiation. A simple
rule has been suggested, whereby the decreased `hidden-variable entropy' $S_{%
\mathrm{hv}}$ (equal to minus the subquantum $H$-function (\ref{H})) of the
outgoing nonequilibrium radiation balances the increase in von Neumann
entropy $S_{\mathrm{vonN}}=-\mathrm{Tr}(\hat{\rho}\ln \hat{\rho})$ generated
by the pure-to-mixed transition:%
\begin{equation}
\Delta \left( S_{\mathrm{hv}}+S_{\mathrm{vonN}}\right) =0\ .  \label{ConsS}
\end{equation}%
Possible experimental tests of this proposal are discussed in refs. \cite%
{AVBHs,AV07}.

It is sometimes suggested that, at the Planck scale, processes will occur
involving the formation and evaporation of microscopic black holes. If one
takes this (rather heuristic) picture seriously and combines it with the
above proposal, one is led to the conclusion that quantum nonequilibrium
will be generated at the Planck scale. During the inflationary phase, such
processes might have an effective description in terms of nonequilibrium
modes of the inflaton field at Planckian or trans-Planckian (physical)
frequencies. One might reason as follows. If a mode of comoving wavelength $%
\lambda $ once had a physical wavelength $\lambda _{\mathrm{phys}}=a\lambda
\lesssim l_{\mathrm{P}}$ near the beginning of inflation, one could assume
that upon exiting the Planckian regime (that is, once $\lambda _{\mathrm{phys%
}}$ becomes bigger than $l_{\mathrm{P}}$) the mode will be out of
equilibrium, having encountered some gravitational process that generates
nonequilibrium while $\lambda _{\mathrm{phys}}\sim l_{\mathrm{P}}$, whereas
modes that were never smaller than $l_{\mathrm{P}}$ will not encounter any
such process. Roughly, one could model this by introducing a cutoff $\lambda
_{\mathrm{c}}^{\prime }$ such that nonequilibrium exists only for comoving
wavelengths $\lambda \lesssim \lambda _{\mathrm{c}}^{\prime }$ (\textit{below%
} the critical value $\lambda _{\mathrm{c}}^{\prime }$, in contrast with the
scenario in the preceding section). In addition to providing an estimate for 
$\lambda _{\mathrm{c}}^{\prime }$, one also needs to estimate the degree of
nonequilibrium, which for a given mode $\mathbf{k}r$ may be quantified by
the relative (or hidden-variable) entropy%
\begin{equation}
S_{\mathrm{hv}}(k)\equiv -\int dq_{\mathbf{k}r}\ \rho _{\mathbf{k}r}\ln
(\rho _{\mathbf{k}r}/\left\vert \psi _{\mathbf{k}r}\right\vert ^{2})\ .
\label{Shvk}
\end{equation}%
An estimate for $S_{\mathrm{hv}}(k)$ might arise from an application of (\ref%
{ConsS}) in some form, though this remains to be studied.

It is to be hoped that further development of this idea will lead to a
detailed prediction for the form and magnitude of nonequilibrium for modes
emerging from the Planckian regime. Pending such development, again, in this
paper we restrict ourselves to using current data to set limits on any
hypothetical quantum nonequilibrium that may be present during the
inflationary phase.

\section{Measuring primordial quantum fluctuations}

The above considerations suggest that, during inflation, some field modes
may exhibit nonequilibrium fluctuations that violate quantum theory. Our aim
in this paper is to show how to use CMB data to set bounds on such
violations.

We shall first recall how measurements of the CMB today allow us to infer
statistical properties of inflaton fluctuations during the inflationary era.
This involves working backwards from the CMB data, first to classical
curvature perturbations in the early universe, and from these, backwards
even further to inflaton fluctuations during the inflationary phase. After
having highlighted the key assumptions that are made in the standard
treatment, we will be in a position to understand exactly how corrections to
quantum theory during inflation are able to have an effect on the CMB.

\subsection{CMB observations and primordial curvature perturbations}

Employing angular coordinates $(\theta ,\phi )$ on the sky, CMB measurements
provide us with a temperature function $T(\theta ,\phi )$. Writing $\Delta
T(\theta ,\phi )\equiv T(\theta ,\phi )-\bar{T}$, where $\bar{T}$ is the
average temperature over the sky, the temperature anisotropy may be
decomposed into spherical harmonics,%
\begin{equation}
\frac{\Delta T(\theta ,\phi )}{\bar{T}}=\sum_{l=2}^{\infty
}\sum_{m=-l}^{+l}a_{lm}Y_{lm}(\theta ,\phi )  \label{har}
\end{equation}%
(where as usual we omit the dipole term). A mode $l$ corresponds to an
angular scale $\approx 60%
{{}^\circ}%
/l$.

A complete measurement of the microwave sky provides us with one function $%
T(\theta ,\phi )$, or equivalently with one set $\left\{ a_{lm}\right\} $ of
coefficients. In order to carry out a statistical analysis of $\left\{
a_{lm}\right\} $, it is usually assumed (if only implicitly) that the
observed $T(\theta ,\phi )$ is a single realisation of a stochastic process,
whose probability distribution $P[T(\theta ,\phi )]$ (which may be thought
of as representing a theoretical `ensemble of skies') satisfies the
condition of statistical isotropy:%
\begin{equation}
P[T(\theta -\delta \theta ,\phi -\delta \phi )]=P[T(\theta ,\phi )]
\label{StatRot}
\end{equation}%
for arbitrary angular displacements $\delta \theta $, $\delta \phi $. This
condition implies that, for a given $l$, each $a_{lm}$ has the same
(marginal) probability distribution $p_{l}(a_{lm})$, with variance%
\begin{equation}
C_{l}\equiv \left\langle \left\vert a_{lm}\right\vert ^{2}\right\rangle
\label{Cl}
\end{equation}%
(the angular power spectrum, where $\left\langle ...\right\rangle $ denotes
an average over the theoretical ensemble).

Thus, given the assumption (\ref{StatRot}), it follows that for each $l$ we
have what are, in effect, $2l+1$ independent realisations of the same random
variable (with the same probability distribution). The observed quantity%
\begin{equation*}
C_{l}^{\mathrm{sky}}\equiv \frac{1}{2l+1}\sum_{m=-l}^{+l}\left\vert
a_{lm}\right\vert ^{2}
\end{equation*}%
(constructed from measurements made on a single sky) then provides an
unbiased estimate of the angular power spectrum $C_{l}$ (that is, $%
\left\langle C_{l}^{\mathrm{sky}}\right\rangle =C_{l}$), with a `cosmic
variance' given by%
\begin{equation}
\frac{\Delta C_{l}^{\mathrm{sky}}}{C_{l}}=\sqrt{\frac{2}{2l+1}}\ .
\label{CV}
\end{equation}%
For large values of $l$, the quantity $C_{l}^{\mathrm{sky}}$ is an accurate
estimate of $C_{l}$ (that is, we expect to find $C_{l}^{\mathrm{sky}}\approx
C_{l}$). For small values of $l$, however, $C_{l}^{\mathrm{sky}}$ is an
inaccurate estimate of $C_{l}$.

The observed CMB anisotropy is caused by classical inhomogeneities on the
last scattering surface, when the CMB photons decoupled (together with
effects taking place afterwards as the CMB photons propagate through space
towards us). These inhomogeneities in turn originate from classical
perturbations that were present at much earlier times. In the long
`primordial' period between $t_{\mathrm{exit}}(k)$ and $t_{\mathrm{enter}%
}(k) $ (during which $k<<Ha$, or $\lambda _{\mathrm{phys}}>>H^{-1}$), the
classical curvature perturbation%
\begin{equation}
\mathcal{R}_{\mathbf{k}}\equiv \frac{1}{4}\left( \frac{a}{k}\right)
^{2}\,^{(3)}R_{\mathbf{k}}
\end{equation}%
is time independent. (Here, $^{(3)}R_{\mathbf{k}}$ is the Fourier component
of the spatial curvature scalar on comoving hypersurfaces, that is, on
hypersurfaces with zero momentum density.) To a good first approximation, we
may ignore gravitational waves, in which case $\mathcal{R}_{\mathbf{k}}$ is
the only independent degree of freedom for the classical primordial
perturbations. In terms of $\mathcal{R}_{\mathbf{k}}$, the $a_{lm}$ may be
expressed as \cite{LR99}%
\begin{equation}
a_{lm}=\frac{i^{l}}{2\pi ^{2}}\int d^{3}\mathbf{k}\ \mathcal{T}(k,l)\mathcal{%
R}_{\mathbf{k}}Y_{lm}(\mathbf{\hat{k}})\ ,  \label{alm}
\end{equation}%
where the transfer function $\mathcal{T}(k,l)$ encodes the astrophysical
processes that generate the temperature anisotropy.

A given primordial curvature perturbation $\mathcal{R}_{\mathbf{k}}$ (for
all $\mathbf{k}$) generates one set $\left\{ a_{lm}\right\} $ of
temperature-anisotropy coefficients. A probability distribution $P[\mathcal{R%
}_{\mathbf{k}}]$ for $\mathcal{R}_{\mathbf{k}}$ will generate a probability
distribution $P[\left\{ a_{lm}\right\} ]$ for $\left\{ a_{lm}\right\} $. If
we make the assumption of statistical homogeneity, that $P[\mathcal{R}_{%
\mathbf{k}}]$ is translationally invariant --- that is, in position space, $%
P[\mathcal{R}(\mathbf{x}-\mathbf{d})]=P[\mathcal{R}(\mathbf{x})]$ for
arbitrary displacements $\mathbf{d}$ --- it follows that $\left\langle 
\mathcal{R}_{\mathbf{k}}\mathcal{R}_{\mathbf{k%
{\acute{}}%
}}\right\rangle e^{-i(\mathbf{k}+\mathbf{k}%
{\acute{}}%
)\cdot \mathbf{d}}=\left\langle \mathcal{R}_{\mathbf{k}}\mathcal{R}_{\mathbf{%
k%
{\acute{}}%
}}\right\rangle $ and so%
\begin{equation}
\left\langle \mathcal{R}_{\mathbf{k}}\mathcal{R}_{\mathbf{k%
{\acute{}}%
}}^{\ast }\right\rangle =\delta _{\mathbf{kk}%
{\acute{}}%
}\left\langle \left\vert \mathcal{R}_{\mathbf{k}}\right\vert
^{2}\right\rangle \ .  \label{RkRk}
\end{equation}

From (\ref{alm}) and (\ref{RkRk}), the angular power spectrum (\ref{Cl}) may
be written as%
\begin{equation}
C_{l}=\frac{1}{2\pi ^{2}}\int_{0}^{\infty }\frac{dk}{k}\ \mathcal{T}^{2}(k,l)%
\mathcal{P}_{\mathcal{R}}(k)\ ,  \label{Cl2}
\end{equation}%
where%
\begin{equation}
\mathcal{P}_{\mathcal{R}}(k)\equiv \frac{4\pi k^{3}}{V}\left\langle
\left\vert \mathcal{R}_{\mathbf{k}}\right\vert ^{2}\right\rangle  \label{PPS}
\end{equation}%
is the primordial power spectrum. We shall assume, as is usually done, that $%
\left\langle \left\vert \mathcal{R}_{\mathbf{k}}\right\vert
^{2}\right\rangle $ is a function of $k$ only.

Current measurements of the CMB show that $\mathcal{P}_{\mathcal{R}%
}(k)\approx \mathrm{const}.$ (an approximately flat or scale-free spectrum) 
\cite{5YrWMAP}.

\subsection{Inflationary slow-roll predictions}

Standard inflation predicts an approximately flat primordial power spectrum $%
\mathcal{P}_{\mathcal{R}}(k)$. Let us briefly review how this comes about.

An approximately homogeneous inflaton field $\phi _{0}(t)+\phi (\mathbf{x}%
,t) $ (where $\phi $ is a small perturbation), with a potential $V$, has an
energy density $\rho \approx \frac{1}{2}\dot{\phi}_{0}^{2}+V(\phi _{0})$. In
the slow-roll approximation, $\rho \approx V(\phi _{0})$ is approximately
constant in time. The Friedmann equation $(\dot{a}/a)^{2}=(8\pi G/3)\rho $
then implies an approximate de Sitter expansion, $a\propto e^{Ht}$, where $H=%
\sqrt{(8\pi G/3)V(\phi _{0})}$. The time evolution of $\phi _{0}$ is given by%
\begin{equation*}
3\frac{\dot{a}}{a}\dot{\phi}_{0}+\frac{dV}{d\phi _{0}}=0
\end{equation*}%
(where in the slow-roll approximation we may neglect the term $\ddot{\phi}%
_{0}$). The flatness conditions for $V$ are $\varepsilon <<1$, $\left\vert
\eta \right\vert <<1$, where%
\begin{equation}
\varepsilon \equiv \frac{1}{16\pi G}\left( \frac{1}{V}\frac{dV}{d\phi _{0}}%
\right) ^{2}\ ,\ \ \ \ \ \eta \equiv \frac{1}{8\pi G}\frac{1}{V}\frac{d^{2}V%
}{d\phi _{0}^{2}}\ .  \label{srpars}
\end{equation}

The primordial perturbations are generated by quantum fluctuations during
the slow roll. As a first approximation, the quantum fluctuations may be
calculated for an eternal de Sitter expansion, and in this approximation one
obtains an exactly scale-free primordial power spectrum. Corrections to this
approximation yield small corrections to the scale-free result.

The quantum theory of primordial perturbations has been developed in great
detail \cite{LL00,LR99,MFB92}. In the slow-roll limit ($\dot{H}\rightarrow 0$%
), with $V$ satisfying the flatness conditions, the inflaton perturbation $%
\phi =\phi (\mathbf{x},t)$ evolves like a free massless field (until at
least a few e-folds after $t_{\mathrm{exit}}(k)$ for the mode $\mathbf{k}$).
The quantised field $\hat{\phi}$ is usually assumed to be in the vacuum
state. One may then use standard quantum field theory to calculate the
probability distribution for the inflaton perturbation $\phi $.

It is usually assumed that, a few Hubble times or e-folds after $t_{\mathrm{%
exit}}(k)$ (that is, in the `late-time limit'), the resulting quantum
probability distribution for $\phi $ may be regarded as a classical
probability distribution over classical perturbations $\phi $. This
assumption has been justified by WKB-type classicality at late times \cite%
{GP85}, by squeezing of the inflationary vacuum state \cite{A94,PS96}, and
by environmental decoherence \cite{KPS98,KP98}. The latter, in particular,
seems to distinguish the field configuration basis (that is, the basis of
eigenstates of the field operator $\hat{\phi}$) as a robust pointer basis,
where the relevant interactions are local in field space \cite{KPS98,KP98}.
The resulting distribution of field configurations is then, for practical
purposes, indistinguishable from a classical distribution. Recent studies
seem to confirm these conclusions: the pointer states consist (more
precisely) of narrow Gaussians that approximate eigenstates of $\hat{\phi}$ 
\cite{KLPS06}, and the locality of interactions in field space ensures that
at late times the density matrix becomes essentially diagonal in the field
configuration basis \cite{BHH06}. (See also ref. \cite{LS06} for further
discussion of WKB classicality in the late-time limit.)

Given a classical inflaton perturbation $\phi $, the corresponding curvature
perturbation is given by \cite{LL00}%
\begin{equation}
\mathcal{R}_{\mathbf{k}}=-\left[ \frac{H}{\dot{\phi}_{0}}\phi _{\mathbf{k}}%
\right] _{t=t_{\ast }(k)}\ ,  \label{Rk}
\end{equation}%
where $t_{\ast }(k)$ is a time a few e-folds after $t_{\mathrm{exit}}(k)$.
The perturbation $\mathcal{R}_{\mathbf{k}}$ is time independent between $%
t_{\ast }(k)$ and the approach to $t_{\mathrm{enter}}(k)$ (long after
inflation ends), and is believed to seed what eventually grow into the
dominant perturbations in the CMB.

Note that the inflaton perturbation $\phi $ is defined on a spatially flat
slicing. (The inhomogeneous field $\phi $ necessarily vanishes on comoving
slices, since the momentum density $-\dot{\phi}\mathbf{\nabla }\phi $ is by
definition zero on such slices.) Then, in the slow-roll limit $\dot{H}%
\rightarrow 0$, the back-reaction of metric perturbations on $\phi $ can be
ignored \cite{LL00}. The curvature perturbation $\mathcal{R}$ is defined on
the comoving slicing. Thus, (\ref{Rk}) relates quantities defined on
different slicings.

The predicted (quantum-theoretical) primordial power spectrum, for $\mathcal{%
R}_{\mathbf{k}}$ at $t=t_{\ast }$, is then given by%
\begin{equation}
\mathcal{P}_{\mathcal{R}}^{\mathrm{QT}}(k)=\left[ \frac{H^{2}}{\dot{\phi}%
_{0}^{2}}\mathcal{P}_{\phi }^{\mathrm{QT}}(k)\right] _{t_{\ast }(k)}\ ,
\label{PRQT1}
\end{equation}%
where%
\begin{equation}
\mathcal{P}_{\phi }^{\mathrm{QT}}(k)\equiv \frac{4\pi k^{3}}{V}\left\langle
\left\vert \phi _{\mathbf{k}}\right\vert ^{2}\right\rangle _{\mathrm{QT}}
\end{equation}%
is the power spectrum of the inflaton fluctuations.

As we have said, to a first approximation the inflaton fluctuations are
usually taken to be quantum vacuum fluctuations in de Sitter spacetime. From
the standard field operator expansion%
\begin{equation}
\hat{\phi}(\mathbf{x},t)=\sum_{\mathbf{k}}\left( \frac{(k/a+iH)}{k\sqrt{2Vk}}%
\hat{a}_{\mathbf{k}}e^{i(\mathbf{k}\cdot \mathbf{x}+k/Ha)}+\frac{(k/a-iH)}{k%
\sqrt{2Vk}}\hat{a}_{\mathbf{k}}^{\dag }e^{-i(\mathbf{k}\cdot \mathbf{x}%
+k/Ha)}\right)
\end{equation}%
in terms of mode functions%
\begin{equation}
\phi _{+}(\mathbf{x},t)\propto \left( \frac{k}{a}+iH\right) e^{i(\mathbf{k}%
\cdot \mathbf{x}+k/Ha)}
\end{equation}%
(solutions of (\ref{weqn}) reducing to positive-frequency Minkowski modes in
the short-wavelength limit $k/a>>H$), the Bunch-Davies vacuum is defined by $%
\hat{a}_{\mathbf{k}}\left\vert 0\right\rangle =0$ (for all $\mathbf{k}$). In
this quantum state, the two-point (equal-time) correlation function is%
\begin{equation}
\langle 0|\hat{\phi}(\mathbf{x},t)\hat{\phi}(\mathbf{x}%
{\acute{}}%
,t)|0\rangle =\sum_{\mathbf{k}}\frac{(k/a)^{2}+H^{2}}{2Vk^{3}}e^{i\mathbf{k}%
\cdot (\mathbf{x}-\mathbf{x}%
{\acute{}}%
)}  \label{2pt}
\end{equation}%
(where the first term in the numerator gives a Minkowskian contribution $%
1/4\pi ^{2}a^{2}|\mathbf{x}-\mathbf{x}%
{\acute{}}%
|^{2}$). The quantum variance of each mode is given by the Fourier transform
of the quantum two-point function,%
\begin{equation*}
\left\langle |\phi _{\mathbf{k}}|^{2}\right\rangle _{\mathrm{QT}}=\frac{V}{%
(2\pi )^{3}}\int d^{3}\mathbf{x}\;e^{-i\mathbf{k}\cdot \mathbf{x}}\langle 0|%
\hat{\phi}(\mathbf{x+y})\hat{\phi}(\mathbf{y})|0\rangle \ ,
\end{equation*}%
yielding%
\begin{equation}
\left\langle |\phi _{\mathbf{k}}|^{2}\right\rangle _{\mathrm{QT}}=\frac{V}{%
2(2\pi )^{3}}\frac{H^{2}}{k^{3}}\left( 1+\frac{k^{2}}{H^{2}a^{2}}\right) \ .
\end{equation}%
The width decreases with time, tending to a finite constant. The power
spectrum is%
\begin{equation}
\mathcal{P}_{\phi }^{\mathrm{QT}}(k)=\frac{k^{2}}{4\pi ^{2}a^{2}}+\frac{H^{2}%
}{4\pi ^{2}}\ .
\end{equation}%
In the long-wavelength limit $k/a<<H$ ($\lambda _{\mathrm{phys}}>>H^{-1}$),
where the mode is well outside the Hubble radius, we have%
\begin{equation}
\mathcal{P}_{\phi }^{\mathrm{QT}}(k)=\frac{H^{2}}{4\pi ^{2}}\ .
\label{phiHZ}
\end{equation}%
(If instead we set $k=Ha$, then $\mathcal{P}_{\phi }^{\mathrm{QT}%
}(k)=H^{2}/2\pi ^{2}$.)

To a lowest-order approximation, then, the quantum fluctuations of the
inflaton field generate a scale-free spectrum of primordial curvature
perturbations:%
\begin{equation}
\mathcal{P}_{\mathcal{R}}^{\mathrm{QT}}(k)=\frac{1}{4\pi ^{2}}\left[ \frac{%
H^{4}}{\dot{\phi}_{0}^{2}}\right] _{t_{\ast }(k)}\ .  \label{PRQT2}
\end{equation}%
These perturbations $\mathcal{R}_{\mathbf{k}}\propto \phi _{\mathbf{k}}$
remain frozen outside the Hubble radius until the time $t_{\mathrm{enter}%
}(k) $ is approached.

Because $H$ and $\dot{\phi}_{0}$ are in fact slowly changing during the
inflationary phase, higher-order corrections lead to a small dependence of $%
\mathcal{P}_{\mathcal{R}}^{\mathrm{QT}}(k)$ on $k$.

\section{Time evolution of nonequilibrium vacua}

We now turn to the effect of quantum nonequilibrium on the predictions of
inflationary cosmology. (A brief, preliminary account was given in ref. \cite%
{AV07}.) Our strategy is to consider nonequilibrium corrections to the
lowest-order (scale-free) quantum spectrum, and then to compare these
effects with the higher-order quantum corrections.

In the Bunch-Davies vacuum, a mode $\mathbf{k}r$ has wave function $\psi _{%
\mathbf{k}r}=\psi _{\mathbf{k}r}(q_{\mathbf{k}r},t)=\left\vert \psi _{%
\mathbf{k}r}\right\vert e^{is_{\mathbf{k}r}}$ with a Gaussian amplitude%
\begin{equation}
\left\vert \psi _{\mathbf{k}r}\right\vert ^{2}=\frac{1}{\sqrt{2\pi \Delta
_{k}^{2}}}e^{-q_{\mathbf{k}r}^{2}/2\Delta _{k}^{2}}  \label{psi2}
\end{equation}%
of width%
\begin{equation}
\Delta _{k}^{2}=\frac{H^{2}}{2k^{3}}\left( 1+\frac{k^{2}}{H^{2}a^{2}}\right)
\label{D2}
\end{equation}%
(contracting in time, and independent of $r$ and of the direction of $%
\mathbf{k}$) and with a phase%
\begin{equation}
s_{\mathbf{k}r}=-\frac{ak^{2}q_{\mathbf{k}r}^{2}}{2H(1+k^{2}/H^{2}a^{2})}%
+h(t)\ ,  \label{s}
\end{equation}%
where%
\begin{equation*}
h(t)=\frac{1}{2}\left( \frac{k}{Ha}-\tan ^{-1}\left( \frac{k}{Ha}\right)
\right)
\end{equation*}%
is independent of $q_{\mathbf{k}r}$. (It is readily verified that the above
wave function $\psi _{\mathbf{k}r}(q_{\mathbf{k}r},t)$ satisfies the Schr%
\"{o}dinger equation (\ref{Sch3}) for a mode $\mathbf{k}r$, and that in the
limit $H\longrightarrow 0$, $a\rightarrow 1$ one recovers the wave function $%
\psi _{\mathbf{k}r}(q_{\mathbf{k}r},t)\propto e^{-kq_{\mathbf{k}r}^{2}}e^{-i%
\frac{1}{2}kt}$ for the Minkowski vacuum.)

In the quantum vacuum, the $q_{\mathbf{k}r}$ are independent random
variables, each with a Gaussian distribution of zero mean. The width of each
Gaussian decreases with time, approaching the asymptotic value $H/\sqrt{%
2k^{3}}$ (in the long-wavelength limit $k/a<<H$). In the nonequilibrium (de
Broglie-Bohm) vacuum, in contrast, each $q_{\mathbf{k}r}$ evolves
deterministically in time, and the probability distribution for each $q_{%
\mathbf{k}r}$ depends on what the probability distribution was at some
`initial' time.

The phase (\ref{s}) implies a de Broglie velocity field%
\begin{equation}
\frac{dq_{\mathbf{k}r}}{dt}=\frac{1}{a^{3}}\frac{\partial s_{\mathbf{k}r}}{%
\partial q_{\mathbf{k}r}}=-\frac{k^{2}Hq_{\mathbf{k}r}}{k^{2}+H^{2}a^{2}}\ .
\label{qdot}
\end{equation}

To solve (\ref{qdot}) for the trajectories $q_{\mathbf{k}r}(t)$, it is
convenient to introduce the conformal time $\eta $, defined by $d\eta =dt/a$%
. (For $a\propto e^{Ht}$ we have $\eta =-1/Ha$; as $t$ runs from $-\infty $
to $+\infty $, $\eta $ runs from $-\infty $ to $0$.) In terms of $\eta $,
the equation of motion for $q_{\mathbf{k}r}$ reads%
\begin{equation}
\frac{dq_{\mathbf{k}r}}{d\eta }=\frac{k^{2}\eta q_{\mathbf{k}r}}{1+k^{2}\eta
^{2}}\ ,  \label{dqdeta}
\end{equation}%
which has the solution%
\begin{equation}
q_{\mathbf{k}r}(\eta )=q_{\mathbf{k}r}(0)\sqrt{1+k^{2}\eta ^{2}}\ .
\label{traj}
\end{equation}%
The width of the packet is given by%
\begin{equation}
\Delta _{k}^{2}=\frac{H^{2}}{2k^{3}}\left( 1+k^{2}\eta ^{2}\right) \ .
\label{D2eta}
\end{equation}

An arbitrary distribution $\rho _{\mathbf{k}r}(q_{\mathbf{k}r},\eta )$
(generally $\neq \left\vert \psi _{\mathbf{k}r}(q_{\mathbf{k}r},\eta
)\right\vert ^{2}$) necessarily satisfies the continuity equation%
\begin{equation*}
\frac{\partial \rho _{\mathbf{k}r}}{\partial \eta }+\frac{\partial }{%
\partial q_{\mathbf{k}r}}\left( \rho _{\mathbf{k}r}\frac{dq_{\mathbf{k}r}}{%
d\eta }\right) =0\ ,
\end{equation*}%
which for the velocity field (\ref{dqdeta}) has the solution%
\begin{equation}
\rho _{\mathbf{k}r}(q_{\mathbf{k}r},\eta )=\frac{1}{\sqrt{1+k^{2}\eta ^{2}}}%
\rho _{\mathbf{k}r}(q_{\mathbf{k}r}/\sqrt{1+k^{2}\eta ^{2}},0)
\label{rhoeta}
\end{equation}%
for any given $\rho _{\mathbf{k}r}(q_{\mathbf{k}r},0)$.

The time evolution amounts to a simple (homogeneous) contraction of both $%
\left\vert \psi _{\mathbf{k}r}\right\vert ^{2}$ and $\rho _{\mathbf{k}r}$.
At times $\eta <0$, $\left\vert \psi _{\mathbf{k}r}\right\vert ^{2}$ is a
contracting Gaussian packet of width $\Delta _{k}(\eta )=\Delta _{k}(0)\sqrt{%
1+k^{2}\eta ^{2}}$, and in the late-time limit $\eta \rightarrow 0$, $%
\left\vert \psi _{\mathbf{k}r}\right\vert ^{2}$ approaches a static Gaussian
of width $\Delta _{k}(0)=H/\sqrt{2k^{3}}$. At times $\eta <0$, $\rho _{%
\mathbf{k}r}$ is a contracting arbitrary distribution of width $D_{\mathbf{k}%
r}(\eta )=D_{\mathbf{k}r}(0)\sqrt{1+k^{2}\eta ^{2}}$ (with arbitrary $D_{%
\mathbf{k}r}(0)$), and in the late-time limit $\eta \rightarrow 0$, $\rho _{%
\mathbf{k}r}$ approaches a static packet of width $D_{\mathbf{k}r}(0)$
(where the asymptotic packet differs from the earlier packet by a
homogeneous rescaling of $q_{\mathbf{k}r}$, as in (\ref{rhoeta})).

For simplicity, we assume that (like $\Delta _{k}$) the nonequilibrium width 
$D_{\mathbf{k}r}$ is independent of $r$ and of the direction of $\mathbf{k}$%
, so that $D_{\mathbf{k}r}=D_{k}(t)$. We then have the result%
\begin{equation}
\frac{D_{k}(t)}{\Delta _{k}(t)}=(\mathrm{const.\ in\ time})\equiv \sqrt{\xi
(k)}\ .
\end{equation}

Note that, for each mode, the `nonequilibrium factor' $\xi (k)$ may be
defined at any convenient fiducial time (in particular, not necessarily at
the same time for every $k$). At least in this lowest-order approximation
for the quantum state $\Psi $, it makes no difference whether we set the
initial conditions for nonequilibrium at the same time for all values of $k$%
, or at different times for different values of $k$ (for example, at $t(k)$
such that $\lambda _{\mathrm{phys}}(k)$ exceeds some critical value).

\section{Nonequilibrium power spectrum}

The above result for the nonequilibrium Bunch-Davies vacuum may be written as%
\begin{equation}
\left\langle |\phi _{\mathbf{k}}|^{2}\right\rangle =\left\langle |\phi _{%
\mathbf{k}}|^{2}\right\rangle _{\mathrm{QT}}\xi (k)\ .
\end{equation}%
This implies that the nonequilibrium power spectrum for the inflaton
fluctuations takes the form%
\begin{equation}
\mathcal{P}_{\phi }(k)=\mathcal{P}_{\phi }^{\mathrm{QT}}(k)\xi (k)\ ,
\end{equation}%
which for $k/a<<H$ reads%
\begin{equation}
\mathcal{P}_{\phi }(k)=\frac{H^{2}}{4\pi ^{2}}\xi (k)\ .  \label{nePS}
\end{equation}

The primordial power spectrum for the curvature perturbations is then%
\begin{equation*}
\mathcal{P}_{\mathcal{R}}(k)=\mathcal{P}_{\mathcal{R}}^{\mathrm{QT}}(k)\xi
(k)\ ,
\end{equation*}%
where $\mathcal{P}_{\mathcal{R}}^{\mathrm{QT}}(k)$ is given by (\ref{PRQT2}%
). Thus we have%
\begin{equation}
\mathcal{P}_{\mathcal{R}}(k)=\frac{\xi (k)}{4\pi ^{2}}\left[ \frac{H^{4}}{%
\dot{\phi}_{0}^{2}}\right] _{t_{\ast }(k)}\ .  \label{PRNE}
\end{equation}

In general, $\xi (k)\neq 1$ and scale invariance is broken. In future work,
along the lines outlined in section 4, we hope to be able to predict
features of the function $\xi (k)$. For the purposes of this paper, $\xi (k)$
is (in principle) an arbitrary function to be constrained by observation.

\section{General remarks}

Before considering how CMB data may be used to constrain the nonequilibrium
function $\xi (k)$, we make some general remarks on the above scenario.

\subsection{Transfer of microscopic nonequilibrium to cosmological scales}

We saw in section 6 that, for each mode $\mathbf{k}$ during the inflationary
phase, the respective widths $D_{k}(t)$ and $\Delta _{k}(t)$ of the
nonequilibrium and equilibrium distributions remain in a fixed ratio $%
D_{k}(t)/\Delta _{k}(t)=\sqrt{\xi (k)}$ over time. This holds in the
approximation where the inflationary phase is treated as an exact de Sitter
expansion. At least to a first approximation, then, we may conclude that
quantum nonequilibrium (if it exists) will not relax during the inflationary
phase, but is instead preserved over time.

Furthermore, because of the exponential expansion of physical wavelengths $%
\lambda _{\mathrm{phys}}$ during inflation, nonequilibrium (if there is any
to start with) will not only be preserved but will also be transferred from
microscopic to macroscopic scales. This `magnification' of the
nonequilibrium lengthscale is particularly striking in the late-time or
large-wavelength limit $\lambda _{\mathrm{phys}}>>H^{-1}$, where the de
Broglie velocity field tends to zero for each mode, $\dot{q}_{\mathbf{k}%
r}\rightarrow 0$. In this limit, which takes effect a few e-foldings after
the mode exits the Hubble radius, both $\rho _{\mathbf{k}r}$ and $\left\vert
\psi _{\mathbf{k}r}\right\vert ^{2}$ become \textit{frozen}. Once this
happens, any difference between $\rho _{\mathbf{k}r}$ and $\left\vert \psi _{%
\mathbf{k}r}\right\vert ^{2}$ is preserved, and is transferred to larger and
larger lengthscales as the physical wavelength $\lambda _{\mathrm{phys}%
}=a(t)(2\pi /k)\propto e^{Ht}$ of the mode gets larger and larger. The
frozen nonequilibrium then exists at a physical lengthscale that grows
exponentially with time, from microscopic to macroscopic scales.

Once inflation has ended, there will be a frozen nonequilibrium distribution
of curvature perturbations $\mathcal{R}_{\mathbf{k}}$ at macroscopic
lengthscales. These perturbations are then transferred to cosmological
lengthscales by the subsequent (post-inflationary) Friedmann expansion.

\subsection{Quantum measurement of the inflaton field}

As we saw in section 5.2, in the standard quantum theory of inflationary
cosmology it is usual to assume that, during inflation, when the physical
wavelength of a mode significantly exceeds the Hubble radius, the
corresponding inflaton perturbation effectively `becomes classical' --- in
the sense that the final quantum probability distribution for inflaton (and
hence curvature) perturbations behaves, to a good approximation, like a
classical probability distribution. As mentioned in section 5.2, various
studies seem to confirm the validity of this assumption \cite%
{GP85,A94,PS96,KPS98,KP98,KLPS06,BHH06,LS06}. In particular, the basis of
eigenstates of the field operator $\hat{\phi}$ (suitably smeared with narrow
Gaussians) seems to act as a robust pointer basis, so that the quantum
distribution of field configurations is, for practical purposes,
indistinguishable from a classical distribution \cite{KPS98,KP98,KLPS06}.

In the pilot-wave formulation of inflationary cosmology, there is a
well-defined inflaton configuration or `beable' (in Bell's terminology \cite%
{Bell87}) at all times, even before Hubble exit. In writing the formulas (%
\ref{nePS}) and (\ref{PRNE}), we have tacitly identified the inflaton beable
after Hubble exit with the `classical' inflaton field after Hubble exit ---
where the latter generates the primordial curvature perturbation via
equation (\ref{Rk}). This identification merits some comment.

Generally speaking, in pilot-wave theory, it is the precise value of the
total beable configuration that (together with the wave function) determines
the outcome of a subsequent quantum measurement. However, as a rule, one
must be cautious about identifying beable values with quantum measurement
values: their relationship must be established on a case-by-case basis,
through analysis of the particular measurement process that is taking place.
As is well known \cite{B52,Bell87}, there are circumstances where a quantum
measurement outcome does not provide a faithful record of the actual prior
value of the beable (in which case the so-called quantum `measurement' is in
fact not a true measurement). For instance, in the pilot-wave theory of
nonrelativistic particles, while the outcome of a quantum position
measurement usually has the same value as the actual particle position prior
to the measurement, for a quantum momentum measurement the outcome usually
does not simply coincide with the prior particle momentum given by de
Broglie's velocity formula. Instead, the quantum momentum outcome depends on
the initial particle position in a way that depends on the details of the
measurement process. Thus, while quantum position measurements are usually
`faithful', quantum momentum measurements are usually not.

Similarly, we expect that in the pilot-wave theory of fields, a quantum
measurement of the field configuration will (usually) provide a faithful
record of the value of the actual field beable appearing in the de
Broglie-Bohm dynamics. For other measurements, however, this simple
identification will not hold: instead, the outcomes will depend on the
initial field beable in a way that depends on the details of the
`measurement' process.

Now, in the case at hand, conventional analysis of the quantum-to-classical
transition during inflation indicates that the environment effects a quantum
measurement of the inflaton field in the basis of field configurations \cite%
{KPS98,KP98,KLPS06,BHH06}. If this is correct, then we are indeed justified
in our above identification of the de Broglie-Bohm inflaton field after
Hubble exit with the classical inflaton field after Hubble exit.

Should the conventional analysis (for some reason) turn out to be incorrect
--- in particular, if the quantum-to-classical transition involves effective
quantum measurements of the inflaton field in a basis different from the
field configuration basis --- then there will be a more complicated
relationship between the de Broglie-Bohm inflaton field and the emergent
classical inflaton field, a relationship that will depend on the details of
the effective measurement process. There would then also be a more
complicated relationship between the nonequilibrium distribution for the
inflaton beable and the nonequilibrium distribution for the primordial
curvature perturbations.

\subsection{Weak dependence on pilot-wave dynamics}

It is worth noting that the above results for the time evolution of
nonequilibrium vacua are only weakly dependent on the details of the de
Broglie-Bohm dynamics. The results are in fact determined by just two
features: (a) there is a field beable $\phi (\mathbf{x},t)$ whose time
evolution is continuous and differentiable, and (b) the dynamics is
`separable', in the sense that for a product quantum state $\Psi \lbrack
\phi ,t]=\prod\limits_{\mathbf{k}r}\psi _{\mathbf{k}r}(q_{\mathbf{k}r},t)$
the velocity of each component $q_{\mathbf{k}r}$ is independent of the other 
$q_{\mathbf{k}r}$'s.

To see this, note that from (b) the evolution reduces to that of a
collection of independent one-dimensional systems. Then, in each
one-dimensional configuration space with coordinate $q_{\mathbf{k}r}$, the
local conservation of quantum equilibrium%
\begin{equation}
\frac{\partial \left\vert \psi _{\mathbf{k}r}\right\vert ^{2}}{\partial t}+%
\frac{\partial (\left\vert \psi _{\mathbf{k}r}\right\vert ^{2}v_{\mathbf{k}%
r})}{\partial q_{\mathbf{k}r}}=0\ ,  \label{1Dcont}
\end{equation}%
for some velocity field $v_{\mathbf{k}r}=v_{\mathbf{k}r}(q_{\mathbf{k}r},t)$%
, uniquely fixes $v_{\mathbf{k}r}$ as%
\begin{equation*}
v_{\mathbf{k}r}(q_{\mathbf{k}r},t)=\frac{1}{\left\vert \psi _{\mathbf{k}%
r}(q_{\mathbf{k}r},t)\right\vert ^{2}}\int_{q_{\mathbf{k}r}}^{\infty }dq_{%
\mathbf{k}r}^{\prime }\frac{\partial \left\vert \psi _{\mathbf{k}r}(q_{%
\mathbf{k}r}^{\prime },t)\right\vert ^{2}}{\partial t}
\end{equation*}%
(assuming that $\left\vert \psi _{\mathbf{k}r}\right\vert ^{2}v_{\mathbf{k}%
r} $ vanishes at infinity), as follows immediately by integrating (\ref%
{1Dcont}) with respect to the coordinate $q_{\mathbf{k}r}$, from some fixed
value $q_{\mathbf{k}r}$ to $\infty $.

Thus, for the case at hand, the assumption of a differentiable and separable
evolution fixes the de Broglie-Bohm velocity field uniquely. Note that this
uniqueness arises only because the system reduces to a collection of
independent one-dimensional systems. It is only in one dimension that the
local conservation of quantum equilibrium fixes the velocity field. In two
or more dimensions, other velocity fields are possible, distinct from that
of de Broglie and Bohm \cite{DG98}.

The conditions (a) and (b) could certainly be violated in other
hidden-variables theories. There might, for example, be no field beable $%
\phi (\mathbf{x},t)$ at all. Also, it is possible to have a pilot-wave-type
theory with a non-separable dynamics \cite{Holl05}. Still, property (a)
might well emerge in some limit from a deeper hidden-variables theory. And
property (b) seems desirable, even if not strictly necessary. In any case,
our point here is to emphasise that (a) and (b) are the only features of
pilot-wave dynamics that really enter into our considerations.

\section{Bound on primordial quantum nonequilibrium}

Let us now illustrate how the available data may be used to constrain the
nonequilibrium function $\xi (k)$ appearing in the result (\ref{PRNE}) for
the primordial power spectrum, where the observed spectrum $\mathcal{P}_{%
\mathcal{R}}(k)=\mathcal{P}_{\mathcal{R}}^{\mathrm{QT}}(k)\xi (k)$ consists
of the usual quantum contributions together with possible nonequilibrium
corrections ($\xi \neq 1$).

It is currently a very active field of research to determine the $k$%
-dependence of the observed spectrum $\mathcal{P}_{\mathcal{R}}(k)$, and to
compare the results with the $k$-dependence of the quantum-theoretical
prediction $\mathcal{P}_{\mathcal{R}}^{\mathrm{QT}}(k)$. It is
straightforward to reinterpret these studies as effectively providing
constraints on the nonequilibrium function $\xi (k)$.

The observed spectrum $\mathcal{P}_{\mathcal{R}}(k)$ is usually
parameterised in terms of the spectral index $n(k)$, defined by%
\begin{equation}
n(k)-1\equiv \frac{d\ln \mathcal{P}_{\mathcal{R}}}{d\ln k}\ ,  \label{nk1}
\end{equation}%
and the running of the spectral index, $n%
{\acute{}}%
(k)\equiv dn/d\ln k$. For $n(k)$ approximately constant, it is convenient to
write the power spectrum in the form%
\begin{equation}
\mathcal{P}_{\mathcal{R}}(k)=\mathcal{P}_{\mathcal{R}}(k_{0})\left( \frac{k}{%
k_{0}}\right) ^{n(k)-1}\ ,  \label{nk2}
\end{equation}%
where $k_{0}$ is some chosen reference or pivot point. (Note that the
definitions (\ref{nk1}), (\ref{nk2}) of $n(k)$ generally agree at $k=k_{0}$
only, and they agree for all $k$ if $dn(k)/dk=0$.) The index $n(k)$ may be
written as a Taylor expansion%
\begin{equation*}
n(k)=n_{0}+\frac{1}{2}\ln \left( \frac{k}{k_{0}}\right) n_{0}^{\prime }+\
...\ ,
\end{equation*}%
where $n_{0}\equiv n(k_{0})$ is the spectral index at $k=k_{0}$, and $%
n_{0}^{\prime }\equiv \left( dn/d\ln k\right) _{0}$ is the running of the
spectral index at $k=k_{0}$.

The observed values of $n(k)$, $n%
{\acute{}}%
(k)$ may be used to set bounds on early quantum nonequilibrium. To
illustrate this, we shall consider a best-fit value of $n_{0}$,%
\begin{equation}
n_{0}=0.960_{-0.013}^{+0.014}  \label{WMAP}
\end{equation}%
at $k_{0}=0.002\ \mathrm{Mpc}^{-1}$ \cite{5YrWMAP}. (Adding nonequilibrium
parameters would of course affect the best-fitting procedure, but the value (%
\ref{WMAP}) suffices here for illustration. A best fitting of nonequilibrium
inflationary models to CMB data is outside the scope of this paper.)

We have $\mathcal{P}_{\mathcal{R}}(k)=\mathcal{P}_{\mathcal{R}}^{\mathrm{QT}%
}(k)\xi (k)$, where $\mathcal{P}_{\mathcal{R}}^{\mathrm{QT}}(k)$ is
predicted by standard inflationary theory. One may adopt the following
parameterisation:%
\begin{equation}
\mathcal{P}_{\mathcal{R}}^{\mathrm{QT}}(k)=\mathcal{P}_{\mathcal{R}}^{%
\mathrm{QT}}(k_{0})\left( \frac{k}{k_{0}}\right) ^{n^{\mathrm{QT}}(k)-1}\ ,
\label{nQT}
\end{equation}%
where $n^{\mathrm{QT}}(k)$ is the usual (quantum-theoretical) spectral
index, and%
\begin{equation}
\xi (k)=\xi (k_{0})\left( \frac{k}{k_{0}}\right) ^{\nu (k)-1}\ ,  \label{nu}
\end{equation}%
where $\nu (k)$ is the `nonequilibrium spectral index'. The observed index
(minus 1) is then a sum%
\begin{equation}
(n-1)=(n^{\mathrm{QT}}-1)+(\nu -1)
\end{equation}%
of contributions from quantum theory and from nonequilibrium corrections.

In the exact limit $\dot{H}\rightarrow 0$, we have $n_{\mathrm{QT}}-1=0$;
and in exact quantum equilibrium, we have $\nu -1=0$. Slow-roll inflation
predicts a small tilt \cite{LL00,LR99}%
\begin{equation}
n^{\mathrm{QT}}(k)-1=-6\varepsilon +2\eta
\end{equation}%
where, in the definitions (\ref{srpars}) of $\varepsilon $ and $\eta $, the
quantities $V$ and $dV/d\phi _{0}$ are evaluated at $t_{\mathrm{exit}}(k)$
(for which $k=aH$).

Defining $\nu _{0}\equiv \nu (k_{0})$ and $n_{0}^{\mathrm{QT}}\equiv n^{%
\mathrm{QT}}(k_{0})$, we obtain a bound for $\left\vert \nu
_{0}-1\right\vert $ on the assumption that $|n_{0}^{\mathrm{QT}}-1|$ is
indeed significantly less than $1$ (as predicted by inflation). Otherwise,
in principle, both $n_{0}^{\mathrm{QT}}-1$ and $\nu _{0}-1$ could be large
--- with comparable magnitudes and opposite signs --- and the observed small
value of their sum $n_{0}-1=-0.04_{-0.013}^{+0.014}$ could be an accident.
We assume here that the observed small value $\left\vert n_{0}-1\right\vert
\lesssim 0.1$ is not due to such a `conspiratorial' cancellation. Then,
roughly, we may write (again at $k_{0}=0.002\ \mathrm{Mpc}^{-1}$)%
\begin{equation}
|n_{0}^{\mathrm{QT}}-1|\lesssim 0.1,\;\;\;\;\left\vert \nu _{0}-1\right\vert
\lesssim 0.1\ .
\end{equation}

The bound $\left\vert \nu _{0}-1\right\vert \lesssim 0.1$ on the
nonequilibrium index may be converted into a bound on the hidden-variable
entropy $S_{\mathrm{hv}}(k)$ --- defined by (\ref{Shvk}) --- for modes with $%
k$ close to $k_{0}=0.002\ \mathrm{Mpc}^{-1}$. (As we have seen, $S_{\mathrm{%
hv}}(k)$ is the relative entropy of $\rho _{\mathbf{k}r}$ with respect to $%
\left\vert \psi _{\mathbf{k}r}\right\vert ^{2}$, and is a natural measure of
the difference between $\rho _{\mathbf{k}r}$ and $\left\vert \psi _{\mathbf{k%
}r}\right\vert ^{2}$.) We have $\xi (k)\equiv D_{k}^{2}/\Delta _{k}^{2}$,
where $D_{k}$ and $\Delta _{k}$ are the widths of $\rho _{\mathbf{k}r}$ and $%
\left\vert \psi _{\mathbf{k}r}\right\vert ^{2}$, respectively. We know that $%
\left\vert \psi _{\mathbf{k}r}\right\vert ^{2}$ is a Gaussian packet, and
that in the late-time limit $\Delta _{k}^{2}=H^{2}/2k^{3}$. For the purposes
of illustration, let us model $\rho _{\mathbf{k}r}$ as a Gaussian (of width $%
D_{k}$). We then have%
\begin{equation}
S_{\mathrm{hv}}(k)=\frac{1}{2}\left( 1-\xi (k)+\ln \xi (k)\right) \ ,
\label{Sxi}
\end{equation}%
with $\xi (k)$ parameterised by (\ref{nu}). If $\nu (k)$ varies slowly, then
close to $k_{0}$ we may write $\nu (k)\approx \nu _{0}$. Taking $\xi
(k_{0})=1$ and assuming that $\left\vert \nu _{0}-1\right\vert $ is small,
we have%
\begin{equation*}
S_{\mathrm{hv}}(k)\approx -\frac{1}{4}(\nu _{0}-1)^{2}\ln ^{2}(k/k_{0})\ .
\end{equation*}%
Restricting ourselves to a range of $k$ close to $k_{0}$, such that $%
\left\vert \ln (k/k_{0})\right\vert \lesssim O(1)$, we then have%
\begin{equation}
\left\vert S_{\mathrm{hv}}(k)\right\vert \lesssim \frac{1}{4}(\nu
_{0}-1)^{2}\lesssim 10^{-2}\ .
\end{equation}

Note that approximate equilibrium in this region (close to $k_{0}$) does not
preclude large departures from equilibrium at much smaller or at much larger
values of $k$.

\section{Possible low-power anomaly at small $l$}

In the low-$l$ region (say $l\lesssim 20$), the angular power spectrum is
dominated by the Sachs-Wolfe effect (resulting from non-uniformities in the
local gravitational potential on the last scattering surface).

In this region, $\mathcal{T}^{2}(k,l)=\pi H_{0}^{4}j_{l}^{2}(2k/H_{0})$ \cite%
{LL00}, where $H_{0}$ is the Hubble constant today, so that (using (\ref{Cl2}%
))%
\begin{equation*}
C_{l}=\frac{H_{0}^{4}}{2\pi }\int_{0}^{\infty }\frac{dk}{k}\
j_{l}^{2}(2k/H_{0})\mathcal{P}_{\mathcal{R}}(k)\ .
\end{equation*}%
For $\mathcal{P}_{\mathcal{R}}(k)=\mathrm{const}.$ we then have%
\begin{equation*}
C_{l}\propto \int_{0}^{\infty }\frac{dk}{k}\ j_{l}^{2}(k)=\frac{1}{2l(l+1)}\
,
\end{equation*}%
so that $l(l+1)C_{l}=\mathrm{const}.$ at low $l$ -- the Sachs-Wolfe plateau
-- as seems to be approximately observed. (The integrated Sachs-Wolfe
effect, taking place along the line of sight, adds a small `rise' at very
small $l$.)

It has been suggested that the data contain anomalously low power at small $%
l $, though this is controversial. If there is such low power, it could of
course be due to some inadequate processing of the data (such as in the
modelling of foregrounds) or to some local astrophysical effect. Otherwise,
the signal could be primordial in origin, reflecting an anomaly in the
underlying spectrum $\mathcal{P}_{\mathcal{R}}(k)$ of curvature
perturbations. In the latter case, the explanation might lie in some
modification of the standard inflationary scenario, or in new physics.

If there is a low-power signal at small $l$ requiring new physics, then
quantum nonequilibrium provides a possible candidate. Taking $\mathcal{P}_{%
\mathcal{R}}(k)=\mathcal{P}_{\mathcal{R}}^{\mathrm{QT}}(k)\xi (k)$, and
assuming (to a first approximation) that $\mathcal{P}_{\mathcal{R}}^{\mathrm{%
QT}}(k)=\mathrm{const}.$, we may write%
\begin{equation}
\frac{C_{l}}{C_{l}^{\mathrm{QT}}}=2l(l+1)\int_{0}^{\infty }\frac{dk}{k}\
j_{l}^{2}(2k/H_{0})\xi (k)\ .  \label{rat}
\end{equation}%
If $\xi (k)=1$ everywhere, then $C_{l}/C_{l}^{\mathrm{QT}}=1$. A low-power
anomaly, $C_{l}<C_{l}^{\mathrm{QT}}$, could be explained by having $\xi
(k)<1 $ in some suitable region of $k$-space. Because the integral is
dominated by the scale $k\approx lH_{0}/2$, a significant drop in $C_{l}$
requires $\xi (k)<1$ for $k$ in this region, that is, $\xi (k)<1$ for
wavelengths $\lambda \approx (4\pi /l)H_{0}^{-1}$ (comparable to today's
Hubble radius).

To have $\xi (k)<1$ for a primordial perturbation mode $\mathbf{k}$ means
that the width $D_{k}$ of the nonequilibrium distribution for the
corresponding inflaton mode is less than the quantum equilibrium width $%
\Delta _{k}$. It is reasonable to expect this, if one accepts the scenario
of section 2, according to which quantum noise arises from statistical
relaxation processes (presumably taking place in the very early universe).
On this view, it is natural to assume that early nonequilibrium would have a
less-than-quantum dispersion, or $\xi (k)<1$ --- as opposed to an early
larger-than-quantum dispersion ($\xi (k)>1$) which, while possible in
principle, seems less natural. Furthermore, we saw in section 4.1 that, in a
supposed pre-inflationary era, relaxation to quantum equilibrium can be
suppressed on large spatial scales, and one expects that at the onset of
inflation nonequilibrium is most likely to have survived at wavelengths $%
\lambda \gtrsim \lambda _{\mathrm{c}}$, where the value of $\lambda _{%
\mathrm{c}}$ remains to be estimated (pending the development of an
appropriate pre-inflationary model). Therefore, it appears that a dip $\xi
(k)<1$ in the power spectrum below some critical wave number $k_{\mathrm{c}%
}=2\pi /\lambda _{\mathrm{c}}$ could be naturally explained in terms of
early quantum nonequilibrium surviving from a very early pre-inflationary
era, though this possibility remains to be developed in detail.

As for the possible production of nonequilibrium in the Planckian regime
(section 4.2), in the absence of a more detailed model we are unable to give
any strong argument for $\xi (k)<1$, as opposed to $\xi (k)>1$, for modes
with $\lambda \lesssim \lambda _{\mathrm{c}}^{\prime }$ (where $\lambda _{%
\mathrm{c}}^{\prime }$ remains to be estimated; see section 4.2). Again
modelling $\rho _{\mathbf{k}r}$ as a Gaussian, the hidden-variable entropy $%
S_{\mathrm{hv}}(k)$ for a single mode is given in terms of $\xi (k)$ by (\ref%
{Sxi}). For a given value of $S_{\mathrm{hv}}(k)$ --- perhaps set by some
application of (\ref{ConsS}) --- equation (\ref{Sxi}) possesses two
solutions for $\xi (k)$, one with $\xi <1$ and one with $\xi >1$. That is,
the same nonequilibrium entropy can be achieved by both a less-than-quantum
and a larger-than-quantum dispersion. On the other hand, from the behaviour
of the function $1-\xi +\ln \xi $, one sees that the solution with $\xi <1$
always has a smaller value of $\left\vert \xi -1\right\vert $ than does the
solution with $\xi >1$; that is, the solution with $\xi <1$ has a dispersion
that is closer to the quantum value. On this (weak) basis, one might suggest
that $\xi <1$ will be preferred. A stronger argument for $\xi <1$ might come
from a detailed understanding of the preservation of information by means of
nonequilibrium noise suppression in the outgoing quantum state of an
evaporating black hole.

In any case, focusing here on the comparison with observation, let us
consider the effect at low $l$ of some simple examples of functions $\xi (k)$%
.

As a first example, motivated by a possible long-wavelength suppression of
relaxation at very early (pre-inflationary) times, we take $\xi (k)=0$ for $%
k<k_{\mathrm{c}}$ and $\xi (k)=1$ for $k>k_{\mathrm{c}}$, where the simple
cutoff is used to model a suppression of quantum noise at wavelengths $%
\lambda >\lambda _{\mathrm{c}}=2\pi /k_{\mathrm{c}}$. We then have%
\begin{equation}
\frac{C_{l}-C_{l}^{\mathrm{QT}}}{C_{l}^{\mathrm{QT}}}=-2l(l+1)\int_{0}^{k_{%
\mathrm{c}}}\frac{dk}{k}\ j_{l}^{2}(2k/H_{0})\ .  \label{Cut1}
\end{equation}%
Again, the dominant scale is $k\approx lH_{0}/2$, and the correction to $%
C_{l}$ will be significant only if the range of integration $(0,k_{\mathrm{c}%
})$ overlaps substantially with this scale --- that is, $k_{\mathrm{c}}$
cannot be much smaller than $lH_{0}/2$.

Note that if, instead, we did take $k_{\mathrm{c}}<<lH_{0}/2$, the
correction to $C_{l}$ would not only be small, it would be unobservable even
in principle, because it would be smaller than the cosmic variance (\ref{CV}%
). For $k<<lH_{0}/2$ we have approximately $j_{l}^{2}(2k/H_{0})\approx
\left( 2^{l}l!/(2l+1)!\right) ^{2}(2k/H_{0})^{2l}$, so that%
\begin{equation*}
\frac{C_{l}-C_{l}^{\mathrm{QT}}}{C_{l}^{\mathrm{QT}}}\approx -(l+1)\left( 
\frac{2^{l}l!}{(2l+1)!}\right) ^{2}\left( \frac{2k_{\mathrm{c}}}{H_{0}}%
\right) ^{2l}\ .
\end{equation*}%
This correction falls off rapidly with increasing $l$, and is very small
even for the lowest values of $l$: for example, even taking $2k_{\mathrm{c}%
}/H_{0}\approx 1$, we find $(C_{4}-C_{4}^{\mathrm{QT}})/C_{4}^{\mathrm{QT}%
}\approx -6\times 10^{-6}$. Because such corrections are much smaller than
the cosmic variance $\Delta C_{l}^{\mathrm{sky}}/C_{l}=\sqrt{2/(2l+1)}$,
they cannot be measured meaningfully, even in principle. To obtain a
measurable effect, the cutoff $k_{\mathrm{c}}$ in (\ref{Cut1}) must not be
small compared to $lH_{0}/2$.

A second example is motivated by the possibility of gravitationally-induced
nonequilibrium at small scales, at wavelengths $\lambda \lesssim \lambda _{%
\mathrm{c}}^{\prime }$. If we assume that the nonequilibrium takes the form
of noise suppression ($\xi <1$), one might model this again with a simple
cutoff, taking $\xi (k)=1$ for $k<k_{\mathrm{c}}^{\prime }$ and $\xi (k)=0$
for $k>k_{\mathrm{c}}^{\prime }$, where $k_{\mathrm{c}}^{\prime }=2\pi
/\lambda _{\mathrm{c}}^{\prime }$. We then have%
\begin{equation}
\frac{C_{l}-C_{l}^{\mathrm{QT}}}{C_{l}^{\mathrm{QT}}}=-2l(l+1)\int_{k_{%
\mathrm{c}}^{\prime }}^{\infty }\frac{dk}{k}\ j_{l}^{2}(2k/H_{0})\ .
\label{Cut2}
\end{equation}%
For a significant effect, the range of integration $(k_{\mathrm{c}}^{\prime
},\infty )$ must again overlap substantially with the dominant region $%
k\approx lH_{0}/2$ --- which now implies that $k_{\mathrm{c}}^{\prime }$
cannot be much larger than $lH_{0}/2$.

As a third example, we consider a power law%
\begin{equation}
\xi (k)=\xi (k_{0})\left( \frac{k}{k_{0}}\right) ^{\nu _{0}-1}  \label{nu0}
\end{equation}%
(with constant index $\nu _{0}$). From (\ref{rat}) we then have%
\begin{equation}
\frac{C_{l}}{C_{l}^{\mathrm{QT}}}=2l(l+1)\xi (k_{0})\left( \frac{H_{0}}{%
2k_{0}}\right) ^{\nu _{0}-1}\int_{0}^{\infty }dx\ j_{l}^{2}(x)x^{\nu
_{0}-2}\ ,  \label{PL}
\end{equation}%
where%
\begin{equation*}
\int_{0}^{\infty }dx\ j_{l}^{2}(x)x^{\nu _{0}-2}=\frac{\sqrt{\pi }}{4}\frac{%
\Gamma \lbrack (3-\nu _{0})/2]\Gamma \lbrack l+(\nu _{0}-1)/2]}{\Gamma
\lbrack (4-\nu _{0})/2]\Gamma \lbrack l+(5-\nu _{0})/2]}\ .
\end{equation*}

Should the existence of a low-power anomaly be confirmed, one might try to
match the anomaly with one of the above nonequilibrium spectra (\ref{Cut1}),
(\ref{Cut2}) or (\ref{PL}).

According to the analysis in ref. \cite{3-YrWMAP}, cutting off the power
below a wave number $k_{\mathrm{c}}\sim 3\times 10^{-4}\ \mathrm{Mpc}^{-1}$
(comparable to the inverse Hubble scale $H_{0}=2.4\times 10^{-4}\ \mathrm{Mpc%
}^{-1}$) slightly improves the fit to the three-year Wilkinson Microwave
Anisotropy Probe (WMAP) data, but the improvement does not seem large enough
to justify any conclusion that such a cutoff really exists. Still, the
possibility of reduced power at large scales is worth exploring, since it
could originate from an early nonequilibrium suppression of quantum noise
(as discussed in section 4.1).

\section{Non-random phases and inter-mode correlations}

So far, we have considered only the angular power spectrum $C_{l}$ of the
microwave sky, and how this could be affected by nonequilibrium corrections
to the primordial (scalar) power spectrum $\mathcal{P}_{\mathcal{R}}(k)$.
Here, we consider how primordial non-Gaussianity could arise from early
quantum nonequilibrium.

The primordial curvature perturbations $\mathcal{R}_{\mathbf{k}}$ are
usually assumed to constitute a Gaussian random field, for which the power
spectrum provides a complete characterisation of the statistical properties.
The phases of Gaussian perturbations are randomly distributed, and there are
no inter-mode correlations.

In standard inflationary scenarios, the Gaussianity of $\mathcal{R}_{\mathbf{%
k}}$ arises directly from the Gaussianity of the quantum vacuum fluctuations
of the inflaton perturbation $\phi _{\mathbf{k}}$. (The Gaussianity of $%
\mathcal{R}_{\mathbf{k}}$ is not, as is sometimes claimed, a mere
consequence of the central limit theorem.) In the quantum Bunch-Davies
vacuum, the inflaton probability distribution at conformal time $\eta $ is
given by%
\begin{equation*}
P^{\mathrm{QT}}[\phi ,\eta ]=\left\vert \Psi \lbrack \phi ,\eta ]\right\vert
^{2}=\prod\limits_{\mathbf{k}r}\left\vert \psi _{\mathbf{k}r}(q_{\mathbf{k}%
r},\eta )\right\vert ^{2}\ ,
\end{equation*}%
where, as we saw in section 6, each $\left\vert \psi _{\mathbf{k}%
r}\right\vert ^{2}$ is a Gaussian of zero mean and width $\Delta
_{k}^{2}=(H^{2}/2k^{3})\left( 1+k^{2}\eta ^{2}\right) $. The two-point
function $\langle 0|\hat{\phi}(\mathbf{x},\eta )\hat{\phi}(\mathbf{x}%
{\acute{}}%
,\eta )|0\rangle $ is given by (\ref{2pt}). The three-point function $%
\langle 0|\hat{\phi}(\mathbf{x},\eta )\hat{\phi}(\mathbf{x}%
{\acute{}}%
,\eta )\hat{\phi}(\mathbf{x}%
{\acute{}}%
{\acute{}}%
,\eta )|0\rangle $ vanishes, as do all odd-point functions. Higher $n$-point
functions (for $n$ even) reduce to sums of products of the two-point
function, as expected for a Gaussian random field. In quantum equilibrium,
then, the generation of primordial curvature perturbations $\mathcal{R}_{%
\mathbf{k}}\propto \phi _{\mathbf{k}}$ by inflaton perturbations is a
Gaussian random process.

However, as a general matter of principle, the primordial perturbations
could be non-Gaussian. And if one considers quantum nonequilibrium for the
inflaton field, there is no particular reason why the nonequilibrium
inflaton fluctuations should be Gaussian.

We have already seen that, in quantum nonequilibrium, the probability
distribution $\rho _{\mathbf{k}r}(q_{\mathbf{k}r},\eta )$ for a single mode
of the inflaton field need not take the quantum Gaussian form (\ref{psi2}).
Simple forms of non-Gaussianity include a non-zero skewness or kurtosis of $%
\rho _{\mathbf{k}r}(q_{\mathbf{k}r},\eta )$ (where the marginal $\rho _{%
\mathbf{k}r}(q_{\mathbf{k}r},\eta )$ for $q_{\mathbf{k}r}$ may, in general,
be obtained from a correlated joint distribution $P[q_{\mathbf{k}r},\eta ]$,
as discussed further below). But non-Gaussianity can take on a wide variety
of forms, and various measures of it have been proposed. Some workers have
reported significant primordial non-Gaussianity in the CMB data \cite{YW08},
while others maintain that the data are consistent with primordial
Gaussianity \cite{5YrWMAP}.

Let us show how quantum nonequilibrium can result in non-random phases and
inter-mode correlations for the primordial perturbations.

The coefficients $a_{lm}=\left\vert a_{lm}\right\vert e^{i\varphi _{lm}}$ in
the spherical harmonic expansion (\ref{har}) are of course generally complex
numbers, and their phases $\varphi _{lm}$ contain a lot of information about
the morphology of the temperature anisotropy $\Delta T(\theta ,\phi )$ (see,
for example, ref. \cite{SC05}). Assuming again that the underlying `ensemble
of skies' is statistically rotationally invariant, the probability
distribution for each $\varphi _{lm}$ must be independent of $m$. For a
fixed value of $l$, we then have $2l+1$ phases $\varphi _{lm}$ with the same
probability distribution $p_{l}(\varphi _{lm})$, and for large $l$ we may
use the measured values of the $\varphi _{lm}$ to probe $p_{l}(\varphi
_{lm}) $. At least to a first approximation, current data are consistent
with $p_{l}(\varphi _{lm})$ being uniform on the unit circle. According to
the basic formula (\ref{alm}), each $a_{lm}$ is a linear combination of all
the curvature perturbation components $\mathcal{R}_{\mathbf{k}}$. And
according to the inflationary result (\ref{Rk}), each $\mathcal{R}_{\mathbf{k%
}}$ is proportional to the late-time inflaton perturbation $\phi _{\mathbf{k}%
}$. Thus, the phase $\varphi _{lm}$ of each $a_{lm}$ is ultimately
determined by the phases $\theta _{\mathbf{k}}$ of all the inflaton
perturbation components $\phi _{\mathbf{k}}=\left\vert \phi _{\mathbf{k}%
}\right\vert e^{i\theta _{\mathbf{k}}}$.

In quantum equilibrium, the inflaton phases $\theta _{\mathbf{k}}$ have a
time-independent distribution $\rho _{\mathbf{k}}^{\mathrm{QT}}(\theta _{%
\mathbf{k}})$ that is uniform on the unit circle:%
\begin{equation*}
\rho _{\mathbf{k}}^{\mathrm{QT}}(\theta _{\mathbf{k}})=\frac{1}{2\pi }\ .
\end{equation*}%
This follows immediately from (\ref{psi2}): the real and imaginary parts of $%
\phi _{\mathbf{k}}=\frac{\sqrt{V}}{(2\pi )^{3/2}}\left( q_{\mathbf{k}1}+iq_{%
\mathbf{k}2}\right) $ have a joint Gaussian distribution $\propto e^{-(q_{%
\mathbf{k}1}^{2}+q_{\mathbf{k}2}^{2})/2\Delta _{k}^{2}}$ that is always
constant on circles centred on the origin in the complex $\phi _{\mathbf{k}}$%
-plane.

In quantum nonequilibrium, the inflaton phases can at some initial
(conformal) time $\eta _{i}$ have an arbitrary distribution $\rho _{\mathbf{k%
}}(\theta _{\mathbf{k}},\eta _{i})$. Will the subsequent time evolution
generate a late-time distribution that tends towards uniformity on the unit
circle? Not in the approximation considered here. The trajectories $q_{%
\mathbf{k}r}(\eta )=q_{\mathbf{k}r}(0)\sqrt{1+k^{2}\eta ^{2}}$ obtained in
section 6 imply that%
\begin{equation*}
\theta _{\mathbf{k}}(\eta )=\tan ^{-1}\left( q_{\mathbf{k}2}(\eta )/q_{%
\mathbf{k}1}(\eta )\right) =\tan ^{-1}\left( q_{\mathbf{k}2}(0)/q_{\mathbf{k}%
1}(0)\right) \ .
\end{equation*}%
Thus, during inflation, the phase $\theta _{\mathbf{k}}$ of each inflaton
mode is static, so that any initial nonequilibrium distribution (with
non-random phases) will remain unchanged over time, $\rho _{\mathbf{k}%
}(\theta _{\mathbf{k}},\eta )=\rho _{\mathbf{k}}(\theta _{\mathbf{k}},\eta
_{i})$ for all values of conformal time $\eta $. (In the complex $\phi _{%
\mathbf{k}}$-plane, the evolution of the joint probability distribution for $%
q_{\mathbf{k}1}$, $q_{\mathbf{k2}}$ amounts to a purely radial contraction
with time, so that the distribution $\rho _{\mathbf{k}}(\theta _{\mathbf{k}%
},\eta )$ of phases is time independent.)

We conclude that the time evolution during the inflationary era does not
scramble the phases of the inflaton perturbations. Any initial
non-uniformity (or non-randomness) in the phase distribution will remain
frozen, all the way to the late-time limit $\eta \rightarrow 0$. It would be
interesting, in future work, to explore how this could affect the phases $%
\varphi _{lm}$ of the measured coefficients $a_{lm}$ in the temperature
anisotropy.

We now consider nonequilibrium inter-mode correlations. In section 6 we
assumed, for simplicity, that the nonequilibrium distribution satisfied the
factorisability condition (\ref{Pprod}), so that the modes were uncorrelated
even in nonequilibrium. However, in principle, correlations among modes are
possible: in quantum nonequilibrium, the inflaton modes can be correlated
even though $\left\vert \Psi \right\vert ^{2}$ (for the Bunch-Davies vacuum)
is a product.

In terms of conformal time $\eta $, an arbitrary correlated joint
distribution $P[q_{\mathbf{k}r},\eta ]$ will evolve according to the
continuity equation%
\begin{equation}
\frac{\partial P}{\partial \eta }+\sum_{\mathbf{k}r}\frac{\partial }{%
\partial q_{\mathbf{k}r}}\left( P\frac{dq_{\mathbf{k}r}}{d\eta }\right) =0\ .
\label{ContP3}
\end{equation}%
Because the wave functional is still that of the Bunch-Davies vacuum, the
velocity field $dq_{\mathbf{k}r}/d\eta $ is still given by (\ref{dqdeta})
and the trajectories in configuration space are still given (mode by mode)
by the result (\ref{traj}). Given the trajectories, the general solution of (%
\ref{ContP3}) may be constructed using the property that $P/\left\vert \Psi
\right\vert ^{2}$ is constant along trajectories (where this follows from
the fact that $P$ and $\left\vert \Psi \right\vert ^{2}$ obey the same
continuity equation). Replacing the labels $\mathbf{k}r$ by a single index $%
n $, we may equate%
\begin{equation*}
\frac{P(q_{1}(0),q_{2}(0),....,q_{n}(0),....,0)}{\left\vert \psi
_{1}(q_{1}(0),0)\right\vert ^{2}\left\vert \psi _{2}(q_{2}(0),0)\right\vert
^{2}....\left\vert \psi _{n}(q_{n}(0),0)\right\vert ^{2}....}
\end{equation*}%
with%
\begin{equation*}
\frac{P(q_{1}(\eta ),q_{2}(\eta ),....,q_{n}(\eta ),....,\eta )}{\left\vert
\psi _{1}(q_{1}(\eta ),\eta )\right\vert ^{2}\left\vert \psi _{2}(q_{2}(\eta
),\eta )\right\vert ^{2}....\left\vert \psi _{n}(q_{n}(\eta ),\eta
)\right\vert ^{2}....}\ .
\end{equation*}%
Using the trajectories $q_{n}(\eta )=q_{n}(0)\sqrt{1+k_{n}^{2}\eta ^{2}}$ and%
\begin{equation*}
\frac{\left\vert \psi _{n}(q_{n},\eta )\right\vert ^{2}}{\left\vert \psi
_{n}(q_{n}/\sqrt{1+k_{n}^{2}\eta ^{2}},0)\right\vert ^{2}}=\frac{1}{\sqrt{%
1+k_{n}^{2}\eta ^{2}}}=\frac{\Delta _{n}(0)}{\Delta _{n}(\eta )}
\end{equation*}%
(where the width $\Delta _{n}(\eta )$ is given by (\ref{D2eta})), we deduce
that%
\begin{equation*}
P(q_{1},q_{2},....,q_{n},....,\eta )=P\left( \frac{\Delta _{1}(0)}{\Delta
_{1}(\eta )}q_{1},\frac{\Delta _{2}(0)}{\Delta _{2}(\eta )}q_{2},....,\frac{%
\Delta _{n}(0)}{\Delta _{n}(\eta )}q_{n},....,0\right) \prod\limits_{n}\frac{%
\Delta _{n}(0)}{\Delta _{n}(\eta )}\ .
\end{equation*}%
This is an exact solution for the evolution of an arbitrary distribution,
expressed in terms of the distribution $P\left(
q_{1},q_{2},....,q_{n},....,0\right) $ at conformal time $\eta =0$.

The possibility of nonequilibrium allows the distribution $P[q_{\mathbf{k}%
r},\eta _{i}]$ at some initial time $\eta _{i}$ to be, in principle,
anything at all. To narrow down the range of possibilities, one might impose
the requirement of statistical homogeneity, $P[\phi (\mathbf{x}-\mathbf{d}%
),\eta _{i}]=P[\phi (\mathbf{x}),\eta _{i}]$ (for arbitrary spatial
displacements $\mathbf{d}$).

Clearly, allowing non-random phases and inter-mode correlations in the
inflationary vacuum opens up a number of novel possibilities. Indeed, the
subject of non-Gaussianity for quantum nonequilibrium states deserves to be
developed in more detail.

\section{Conclusion}

We have shown how inflationary cosmology (assuming it to be essentially
correct) may be used to test the validity of quantum theory at very short
distances and at very early times. In particular, we have considered the
possible effects of quantum nonequilibrium, as described by the
hidden-variables theory of de Broglie and Bohm, during the inflationary
phase. We have shown, by means of simple examples, how CMB data may be used
to set bounds on nonequilibrium deviations from quantum theory.

As for the possible origin of such deviations, we have outlined a scenario
where quantum nonequilibrium during the inflationary phase arises from
relaxation suppression (for long-wavelength modes) in a pre-inflationary
era. This scenario suggests that primordial nonequilibrium could set in
above some infra-red cutoff $\lambda _{\mathrm{c}}$ (though the value of $%
\lambda _{\mathrm{c}}$ remains to be estimated). We have also considered the
more speculative possibility that nonequilibrium could be generated during
the inflationary era, by novel gravitational effects at the Planck scale.

We have, for the most part, discussed quantum nonequilibrium corrections to
the primordial (scalar) power spectrum $\mathcal{P}_{\mathcal{R}}(k)$. A
preliminary discussion was also given showing how primordial non-Gaussianity
(in particular, non-random phases and inter-mode correlations) could also
arise from early quantum nonequilibrium.

In this paper we have, for simplicity, considered only the (dominant) scalar
part of the primordial perturbations. The standard quantum theory of
perturbations around a classical background includes tensor contributions
(transverse-traceless metric perturbations, or gravitational waves), as well
as the scalar part considered here \cite{LL00,LR99,MFB92}. It would be
straightforward to extend the present treatment to include tensor
perturbations. The standard formalism may be written in the functional Schr%
\"{o}dinger picture and converted into a de Broglie-Bohm theory in the usual
way, by reinterpreting the quantum probability current in configuration
space in terms of an equilibrium ensemble of trajectories. (As mentioned in
section 2, a de Broglie-Bohm velocity field (\ref{deBgen}) may be defined by
this means for any system with a Hamiltonian given by a differential
operator on configuration space \cite{SV09}.) Once the velocity field for
the trajectories has been identified, one can consider the evolution of an
arbitrary nonequilibrium ensemble. The extended configuration would now
include the transverse-traceless metric perturbations (with two independent
components, corresponding to the two possible states of polarization, each
with approximately the same action as a free massless scalar field). A de
Broglie-Bohm velocity field would be defined for these degrees of freedom as
well.

An important topic that should be examined is how quantum nonequilibrium
would affect the consistency relation between the power spectra for the
scalar and tensor perturbations. This relation is especially interesting
because it relates fluctuations for distinct degrees of freedom, and because
it is independent of the form of the inflaton potential. Presumably, quantum
nonequilibrium would in general have different effects on different degrees
of freedom, resulting in a violation of the consistency relation.

If standard inflationary cosmology is essentially correct, then observations
of the CMB have already confirmed --- to a first approximation --- the
validity of the quantum (Born-rule) prediction for the inflaton power
spectrum during the inflationary phase. More accurate measurements of the
primordial power spectrum will enable us to set unprecedented bounds on
violations of quantum theory, at very short distances and at very early
times. And close scrutiny of other possible features, such as various forms
of non-Gaussianity, will provide further tests of basic quantum predictions.

Should inflation be very firmly established, and should it be found that the
predictions of quantum theory continue to hold well at all accessible
lengthscales during the inflationary era, then this would constitute
considerable evidence against the hypothesis of quantum nonequilibrium at
the big bang (though of course, nonequilibrium from an earlier era might
simply have not survived into the inflationary phase). Furthermore, it would
rather undermine the view that quantum theory is merely an effective
description of an equilibrium state. In principle, one could still believe
that hidden variables exist, and that the hidden-variables distribution is
restricted to quantum equilibrium even at the shortest distances and
earliest times. But in the complete absence of nonequilibrium, the detailed
behaviour of the hidden variables (such as the precise form of the
trajectories in de Broglie-Bohm theory) would be forever untestable. While
exact equilibrium always and everywhere may constitute a logically possible
world, from a general scientific point of view it seems unacceptable, and
the complete ruling out of quantum nonequilibrium by experiment would
suggest that hidden-variables theories should be abandoned.

On the other hand, a positive detection of quantum nonequilibrium phenomena
in the early universe (or indeed elsewhere \cite{AV07}) would be of
fundamental interest, opening up a new and deeper level of nature to
experimental investigation.

\textbf{Acknowledgements.} This work was partly supported by Grant No.
RFP1-06-13A from The Foundational Questions Institute (fqxi.org). For their
hospitality, I am grateful to Carlo Rovelli and Marc Knecht at the Centre de
Physique Th\'{e}orique (Luminy), to Susan and Steffen Kyhl in Cassis, and to
Jonathan Halliwell at Imperial College London.

\end{document}